\documentclass[aps,pra,twocolumn,groupedaddress,floatfix]{revtex4-1}
\usepackage{amsmath}
\usepackage{amssymb}
\usepackage{epsfig}
\usepackage{tikz}
\usepackage{dsfont}
\usepackage{url}
\usepackage[colorlinks]{hyperref}
\hypersetup{%
	plainpages=true,
	breaklinks=true,
	hypertexnames=false,
	pageanchor=true,
	colorlinks=true,
	linkcolor={blue},
	citecolor={red},
	urlcolor={blue},
	anchorcolor={black}
}
\usepackage{bbm}
\usepackage[draft,inline,nomargin,index,marginclue]{fixme}
\fxsetup{status=draft}
\graphicspath{{.}{./figs/}}
\newcommand{\ket}[1]{{\vert #1 \rangle}}

\newcommand{\proj}[1]{{\vert #1 \rangle\langle #1\vert}}

\newcommand* {\bra}[1]{\ensuremath{\langle {#1} |}}
\newcommand{\imag}{\textrm{i}}

\newcommand{\ave}[1]{{\langle #1\rangle}}

\begin{document}
\title{Distant emitters in ultrastrong waveguide QED:\\ Ground-state properties and non-Markovian dynamics}
\author{Carlos A. Gonz\'alez-Guti\'errez}
\email{carlosgg@unizar.es}
\author{Juan Rom\'an-Roche}
\author{David Zueco}
\affiliation{Instituto de Nanociencia y Materiales de Arag\'on (INMA), CSIC-Universidad de Zaragoza, Zaragoza 50009, Spain}
\date{\today}
\begin{abstract}
Starting from the paradigmatic spin-boson model (SBM), we investigate the static and dynamical properties of a system of two distant two-level emitters coupled to a one-dimensional Ohmic waveguide beyond the rotating wave approximation (RWA). Employing static and dynamical polaron \emph{ansätze} we study the effects of finite separation distance on the behavior of the photon-mediated Ising-like interaction, qubit frequency renormalization, ground state magnetization, and entanglement entropy of the two qubit system. 
Based on previous works we derive an effective approximate Hamiltonian for the two-impurity SBM that preserves the excitation-number and thus facilitates the analytical treatment. In particular, it allows us to introduce non-Markovianity arising from delay-feedback effects in two distant emitters in the so-called ultrastrong coupling (USC) regime. We test our results with numerical simulations performed over a discretized circuit-QED model, finding perfect agreement with previous results, and showing interesting new dynamical effects arising in ultrastrong waveguide-QED with distant emitters.
In particular, we revisit the Fermi two-atom problem showing that, in the USC regime, initial correlations yield two different evolutions for symmetric and antisymmetric states even before the emitters become causally connected. 
Finally, we demonstrate that the collective dynamics,  e.g., superradiance or subradiance, are affected not only by the distance between emitters, but also by the coupling, due to significant frequency renormalization. This constitutes another dynamical consequence of the USC regime.
\end{abstract}
\maketitle
\section{Introduction}
\label{intro}
In most scenarios, photons are weakly coupled to matter and they travel fast relative to the matter dimensions.
In waveguide QED (wQED), however, they are confined in one dimensional waveguides to enhance the light-matter coupling \cite{roy2017, gu2017, SolanoReview}.  
Besides, well-separated emitters can be coupled to the guide, such that photons take an appreciable amount of time traveling between them. 
In these cases and depending on the experimental conditions and architectures \cite{SolanoReview,astafiev2010,van2013,faez2014,lodahl2015, liu2017,chang2018,kockum2019, forn2019}, neglecting strong light-matter correlations and/or retardation effects is not always justified.
%
%
When taken into account,  emitter dynamics becomes non-Markovian,   {\it  i.e.}, the system exhibits {\it memory}.
This is interpreted as a back-flow of information between the traveling photons and the emitters.
Non-Markovianity occurs already at the level of a single emitter interacting with  structured  environments or reservoirs \cite{Sajeev1994,vats1998,deVega2017,GonzalezBallestero2013,GonzalezTudelaPRL2017,GonzalezTudela2017,Bello2019, Zueco2019}, in the presence of bound states \cite{sanchez2017b}, or in the ultrastrong coupling (USC) regime \cite{kockum2019, forn2019}.
In this particular regime, higher order processes beyond the creation (annihilation) of \emph{one} photon by annihilating (creating) \emph{one} matter excitation, play a fundamental role.
Entering the USC regime implies that the Rotating-Wave-Approximation (RWA) for the interaction breaks down and the atomic bare parameters get renormalized, either by the Bloch-Siegert shift  \cite{shirley1965},the effective qubit-qubit couplings \cite{zueco2009, Stassi2020} in cavity-QED, or by renormalization  due to the coupling to the continuum electromagnetic (EM) field in wQED \cite{Leggett87}. Besides, the ground state becomes nontrivial \cite{Ashhab2010}.
Apart from non-Markovianity, USC  in wQED has interesting consequences, such as the localization-delocalization transition \cite{peropadre2013, Shi2018},  particle production \cite{gheeraert2018},  non-linear optics at the single photon limit \cite{sanchez2014, sanchez2015} or vacuum light emission \cite{sanchez2019}.

In this work, we will consider this regime of coupling.
When several emitters are present, radiation retardation effects act as an alternative source of non-Markovianity.
This has been recently studied outside of the USC regime showing that even in a simple system of two separated two-level emitters, the collective dynamics can be significantly affected by non-Markovian interference caused by radiation-delayed feedback between them \cite{Baranger2013, Baranger2014,Laakso2014, DiazCamacho2015,Kanu19, Dinc2019,DincPRA2020}. 
%
In this case retardation effects become important when the atomic lifetime $\gamma^{-1}\sim x/v_{g}$,
for a separation distance $x$. In the USC regime this condition depends on the coupling and frequency renormalization, as we will see in the following sections.
%
The case of three distant emitters has also been reported recently in \cite{Regidor2021}.
This retarded back-action can lead to new collective states in which the emission rate of photons by the emitters can be enhanced or inhibited beyond the usual Markovian limit. This effect, termed {\it superduperradiance} \cite{Kanu19}, allows decay rates larger than $2\gamma$,
for two emitters with individual decay rate $\gamma$, which has been predicted to scale linearly with the number of qubits and numerically confirmed up to 100 emitters coupled to a one-dimensional waveguide \cite{Dinc2019}. Non-Markovian features in the emission spectrum of a driven two-qubit system were also investigated recently, together with an experimental proposal in circuit-QED using transmon qubits and Josephson-Junction arrays \cite{Kanu2020}. Another interesting proposal for observing modification of collective phenomena has been laid out in \cite{Longhi2020}, by emulating the emitter dynamics in optical
waveguide arrays. 
In general, memory effects can be viewed as a problem of quantum feedback, where the system is fed with a quantum signal after some time-delay. A general theory of quantum feedback using tensor-networks was developed in \cite{Grissmo2015}, and latter generalized for multiple delays in \cite{wall2017}. Efficient numerical methods to treat more complex systems have also been proposed in \cite{Pichler2016, Regidor2020}, and  the capabilities of feedback to generate universal states for quantum computing was demonstrated in \cite{Zoller17}.

So far, all previous works rely on neglecting counter-rotating terms in the light-matter coupling.
In this work we present a first step in generalizing the effect of distance to the USC coupling regime, combining both sources of non-Markovianity. This is done by considering an Ohmic environment with non-flat spectral density, and by taking into account delay-memory effects due to finite distance separation between emitters. 
Using the polaron transformation, a method that has been shown to be useful for discussing the dynamics of wQED in the USC regime, we first review the ground state properties of two distant emitters as done in \cite{McCutcheon2010}.  
We introduce a discrete model to support our numerical calculations, matching the Ohmic spectral density used throughout the paper. 
We discuss  and characterize the localization-delocalization quantum phase transition, finding closed formulas for the entanglement entropy.
Then, we discuss the dynamics of the emitters in the low energy sector.  A detailed discussion of the Fermi two-atom problem in the USC regime is also given.  Besides, we compute the effective decay rates for initially correlated symmetric (antisymmetric) initial states.
In short, we present a formalism for the study of the different sources of non Markovianity with various emitters in the non perturbative regime of wQED.

The rest of the paper is organized as follows.  The next section deals with the model, the polaron formalism and introduces the discrete model for the waveguide.  Then, the ground state is discussed.  Section \ref{sect:dyn} is the main section of this work.  We discuss the low energy sector of the dynamics.  The paper ends with the conclusions, while several technical aspects are sent to the appendices.


\begin{figure}
\begin{tikzpicture}
    \node [anchor=north west] (imgA) at (0.000\linewidth,.600\linewidth){\includegraphics[width=.35\linewidth]{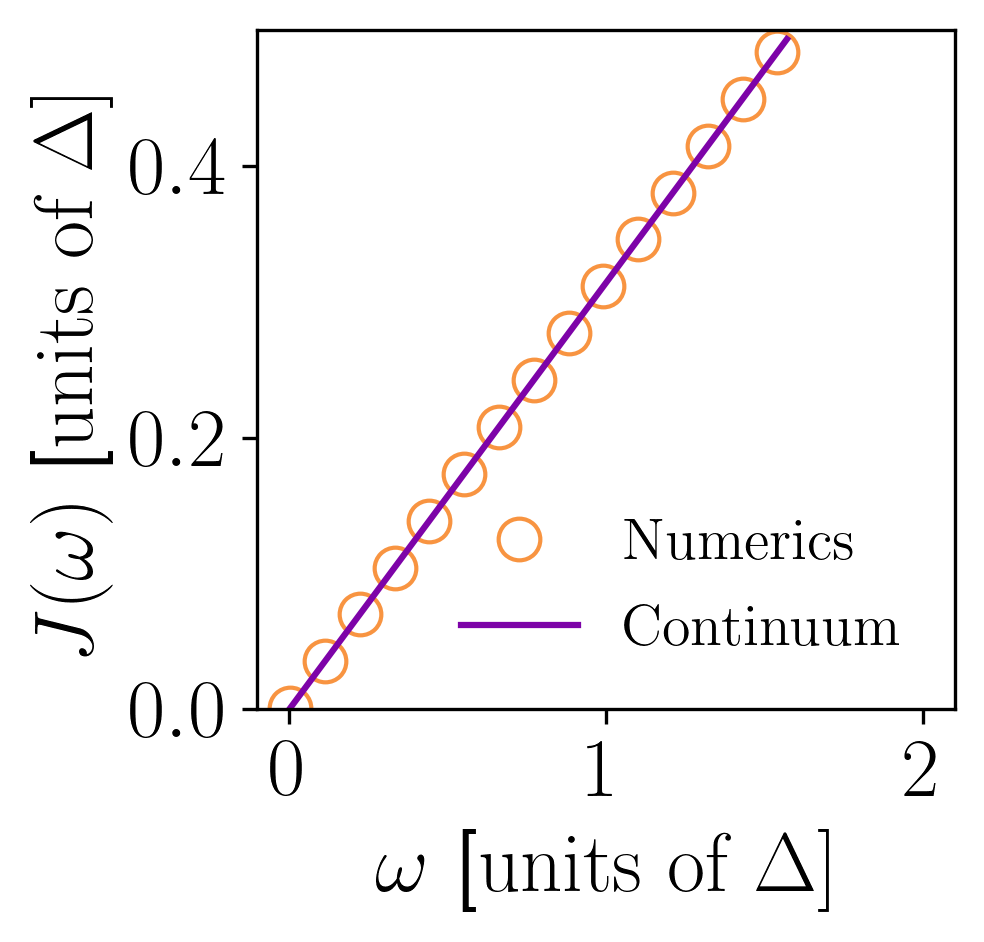}};
    \node [anchor=north west] (imgB) at (0.37\linewidth,.600\linewidth){\includegraphics[width=.62\linewidth]{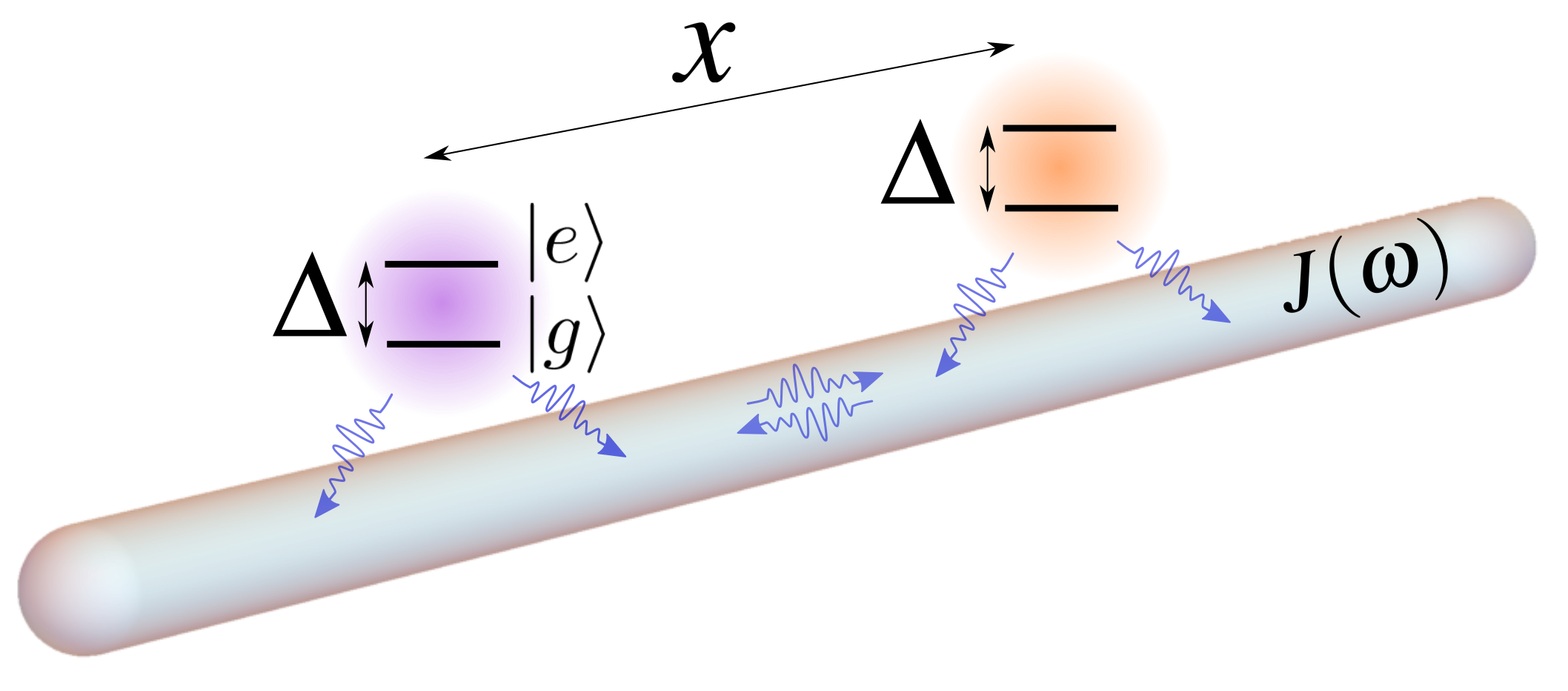}};
    \draw [anchor=north west] (0.000\linewidth, .62\linewidth) node {$\mathsf{(a)}$};
    \draw [anchor=north west] (0.4\linewidth, .62\linewidth) node {$\mathsf{(b)}$};
    \node [anchor=north west] (imgC) at (0.05\linewidth,.200\linewidth){\includegraphics[scale=0.25]{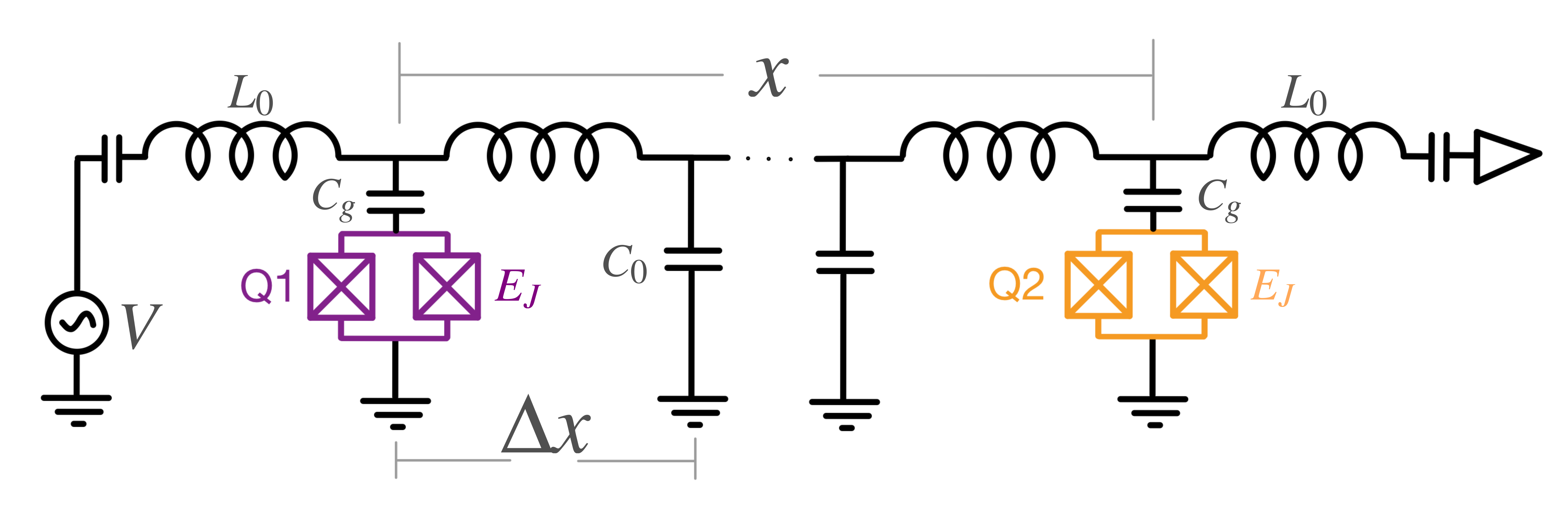}};
    \draw [anchor=north west] (0.\linewidth, .20\linewidth) node {$\mathsf{(c)}$};
\end{tikzpicture}
  \caption{a) Spectral function of the Ohmic waveguide $J(\omega)$. Open circles are obtained from the discrete model simulations with parameters: $\left\{N,\omega_c,\alpha\right\}$ $=\left\{1001,1.0,0.1\right\}$ and the solid line with the formula $J(\omega)=\pi\alpha\omega.$  b) Pictorial illustration of two two-level emitters separated a distance $x$ interacting with the guided modes of an Ohmic one-dimensional waveguide. c) Circuit-QED implementation of the system (see appendix \ref{circuit::spin::boson} for further details).}
  \label{fig1}
\end{figure}
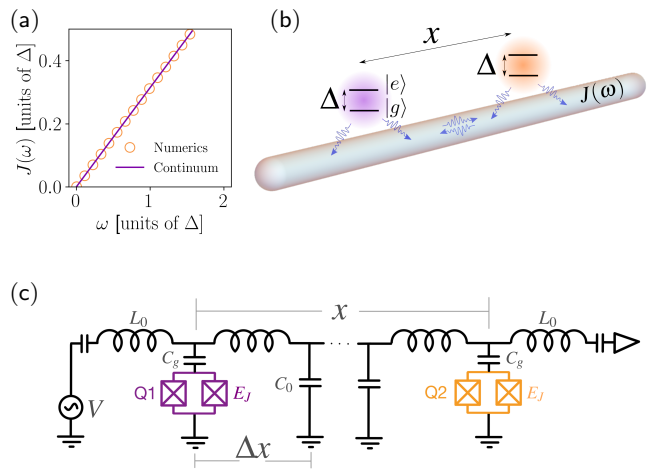
\section{Theoretical Model}
\label{theoretical::model}
We start reviewing the spin-boson model (SBM) for $N_q$  spin-1/2 particles (or qubits) located at positions $x_j$ in one spatial dimension. This model is also known as the $N_q$-impurity SBM and it is considered a paradigmatic Hamiltonian for the understanding of decoherence, dissipation and the physics of open quantum systems \cite{Leggett87,weiss2012quantum}.  Applications of this model include problems from fundamental physics such as the quantum to classical transition \cite{weiss2012quantum}, electronic transport in biological complexes \cite{Xui1994}, and quantum simulators \cite{Lemmer2018}.  The general SBM is described by the following Hamiltonian ($\hbar=1$)
\begin{eqnarray}
\label{spin::boson}
    H=\frac{\Delta}{2}\sum_{j=1}^{N_q}\sigma^{z}_{j}&+&\sum_{k}\omega_k a^{\dagger}_{k}a_{k}\nonumber\\
    &+&\sum_{j=1}^{N_{q}} \sigma^{x}_{j}\sum_{k} g_{k}\left(a_{k}e^{-\imag kx_j}+\rm{h.c.}\right),
\end{eqnarray}
where $\sigma^{z}_{j}=\ket{e}\bra{e}_{j}-\ket{g}\bra{g}_{j}$, $\sigma^{x}_{j}=\ket{e}\bra{g}_{j}+\ket{g}\bra{e}_{j}.$
The SBM was originally introduced for a single two-sate system as a fundamental physical model for decoherence \cite{Leggett87}. It describes 
the interaction of $N_q$ spin-1/2 particles with the surrounding environment considered as a collection of harmonic oscillators at finite temperature (heat bath). 
In wQED, it is the standard model for either actual spin-$1/2$ systems or quantum emitters with anharmonic spectrum after performing the two-level approximation.  
In the latter case, the two level approximation can be questioned at enough strong couplings \cite{DiStefano2019,DeBernardis2018}. 
%
In this work, we fix our attention on natural or artificial two-level emitters, for which the SBM provides an excellent description in the USC regime, our main interest here.
This ultrastrong coupling regime of light and matter has been recently achieved in several experimental systems involving superconducting circuits \cite{Niemczyk2010,forn2017}, semiconductors \cite{Geiser2012}, organic aggregates \cite{Gambino2014}, optomechanical systems \cite{Benz2016}, and more \cite{forn2019,Makihara2021}.

The system-environment coupling can be completely encoded in the so called spectral function of the environment,  defined as \cite{weiss2012quantum}
\begin{eqnarray}
J(\omega)=2\pi\sum_{k}|g_{k}|^2\delta(\omega-\omega_k),
\end{eqnarray}
where $g_k$ is the coupling strength to the $k$-\rm{th} mode of frequency $\omega_k$. The explicit form of this spectral function depends on the physical realization of the corresponding bath. In this paper we will be interested in modeling the environment as an Ohmic waveguide for which $g_k\propto \sqrt{\omega_k}$, giving rise to an Ohmic spectral function $J(\omega)=\pi\alpha\omega$ in the continuum limit, being $\alpha$ the coupling strength parameter (see appendix \ref{discrete::model}). For the single SBM ($N_q=1$) and zero temperature it is well known that, as the coupling increases, there exists a critical value where the system suffers a quantum phase transition called localization transition with vanishing spin magnetization $\ave{\sigma^{z}}=0$  \cite{Chakravarty1982}.  
%
Experimental realizations of the SBM through the coupling of a flux qubit to an open 1D transmission line using a shared Josephson junction have been achieved in Ref. \cite{forn2017}.
In the case of strong coupling, the spontaneous emission rate from the qubit to the waveguide was measured to be $\gamma\sim 2\pi\times88\,\rm{MHz}$ for a qubit frequency of $\Delta\sim 2\pi\times3.99\,\rm{GHz}$. From the microscopic description of the SBM we know that $\alpha=\gamma/\pi\Delta$ (see appendix \ref{alphaOhmic}), which results in a value of $\alpha\approx7\times 10^{-3}.$
In this regime results are well described within the RWA.
On the other hand, to explore the USC regime we require that $\gamma\sim\Delta$, which was also achieved in the same experiment (with a different device), measuring values of $\gamma\sim 2\pi\times9.24\,\rm{GHz}$, and $\Delta\sim 2\pi\times7.68\,\rm{GHz}$, from which we extract the spin boson coupling $\alpha\approx0.38$. This is a clear manifestation that the system has entered the nonperturbative USC regime \cite{forn2017}. From a general view, it seems that experiments with open transmission lines (propagating photons) can explore a wide range of coupling regimes in regards to the SBM, ranging from the underdamped ($\alpha<0.5$) to the localized regime ($\alpha>1.0$). Another recent experiment in the context of the driven SBM \cite{Magazzu2018} has shown that values of $\alpha\sim0.8$ are within the current reach.
%
\\

In general, the SBM can not be diagonalized, but a useful approach based on the introduction of a variational displaced oscillator basis can be employed for the study of static and dynamical properties  of such systems \cite{Silbey&Harris84,McCutcheon2010,DiazCamacho2016,Zueco2019,sanchez2019,JRR2020}. 
%
The accuracy of the polaron approach has been tested in several works for one qubit \cite{Lu&Zheng07,Lee2012,Yao2013,Shi2018}, and multiqubit systems \cite{kurcz2014,Blais2020}.
%
The variational approach is based on the following multiqubit {\it polaron} transformation 
\begin{eqnarray}
\label{polaron:unitary}
    U_P[\{f_{k}\}]&=&\exp\left[-\sum_{j=1}^{N_{q}}\sigma^{x}_{j}\sum_{k}\left(f_{k}a^{\dagger}_{k}e^{\imag k x_{j}}-f^{*}_{k}a_{k}e^{-\imag k x_{j}}\right)\right], \nonumber\\
    &=&\bigotimes_{j=1}^{N_q} U_{j},
\end{eqnarray}
where $U_{j}=\exp[-\sigma^{x}_{j}\sum_k(f_{k}a^{\dagger}_{k}e^{\imag k x_{j}}-\rm{h.c.})]$. The factorization in Eq. (\ref{polaron:unitary}) in local operators acting on the qubits can be done once isotropic propagation of bosons (photons) is assumed, i.e., $|f_{k}|=|f_{-k}|$.
The ground state {\it ansatz} can be defined as the application of the unitary in Eq. (\ref{polaron:unitary}) to a non-entangled state of the qubits and photons
\begin{eqnarray}
\label{ground::state::ansatz}
\ket{\Psi_{\rm GS}
[f_k,\zeta_s]}=
U_{P}\sum_{s_j}\zeta_{s}\ket{s_1,..., s_{N_{q}}}\otimes\ket{\mathbf{0}},
\end{eqnarray}
Where $\ket{s_1,...,s_{N_q}}$ is an arbitrary sate of the qubits and $\ket{\bf{0}}=\ket{0,...,0}$ is the multivacuum state of the photonic waveguide. 
Within this approach the task of finding the ground state energy of the SB system is completely equivalent to minimizing the energy over the following effective $N_q$-qubit Hamiltonian
\begin{eqnarray}
\label{spin::hamiltonian}
H_{S}=\frac{\Delta_r}{2}\sum_{j=1}^{N_q}\sigma^{z}_{j}-\sum_{i<j}\mathcal{J}_{ij}\sigma^{x}_{i}\sigma^{x}_j+2\sum_{k}f_{k}(\omega_k f_k-2g_k), \nonumber\\
\end{eqnarray}
where the qubit frequencies become renormalized according to the rule
\begin{equation}
\label{deltar}
\Delta_r=\Delta\exp\left[-2\sum_{k}|f_{k}|^2\right],
\end{equation}
and the photon-mediated Ising interaction
\begin{eqnarray}
\label{Ising::ij}
\mathcal{J}_{ij}=2\sum_k
f_{k}\left[2g_{k}-\omega_k f_k\right]\cos\left[k(x_{i}-x_{j)}\right].
\end{eqnarray}
The Hamiltonian in Eq. (\ref{spin::hamiltonian}) resembles the one describing the Ising model in the presence of a transverse magnetic field.
In the case of a single-qubit, the ground-state energy in the polaron basis is $E_{GS}=-\Delta_r/2+\sum_{k}f_{k}\left[\omega_{k} f_{k}-2g_{k}\right]$, which can be minimized with respect to the variational parameters $\{f_k\}$. The corresponding free-energy minimization results in a self-consistent relation for the variational parameters \cite{Silbey&Harris84}
\begin{eqnarray}
\label{f_k::one::qubit}
f_{k}=\frac{g_{k}}{\omega_k+\Delta_r}.
\end{eqnarray}
In the case of a single qubit, the  renormalization frequency can be computed explicitly, obtaining  the well known formula in the scaling limit
$\Delta/\omega_c\ll 1$ 
\cite{Leggett87,weiss2012quantum},
\begin{eqnarray}
\label{deltar::onequbit}
\Delta_r\approx\Delta\left(\frac{\Delta}{\omega_c}\right)^{\alpha/(1-\alpha)}.
\end{eqnarray}
It is worth recalling that this renormalization is responsible for the   localization-delocalization phase transition  that corresponds to  the ferromagnetic-antiferromagnetic phase transition in the Kondo model \cite{guinea1998}.
Here, $\Delta_r$  drops to zero as the coupling increases, triggering the  transition at the critical coupling $\alpha_c=1$.
\subsection{Discrete microscopic  model}
Below, we will face situations, mainly when we discuss the dynamics of the system, for which an
analytical solution is not  possible. Then, we approximate the environment using a finite number of modes $N$ by taking advantage of the discrete model for a transmission line resonator exposed in the appendix \ref{discrete::model}. In the continuum limit ($N\rightarrow\infty$) this model reproduces the spectral linear function of an Ohmic environment with an appropriate cutoff frequency: see  Fig. \ref{fig1}(a)  where the points are the spectral function constructed from the discrete model,  recovering the linear dispersion relation $\omega_{k}=v_{g}|k|$. 
This finite $N$-model has already been successfully used to test the dynamics of few emitters interacting with waveguides within \cite{DiazCamacho2015}, and beyond \cite{DiazCamacho2016} the RWA, with numerical methods based on matrix-product-state simulations (MPS), showing remarkable agreement with the advantage of low computational cost. 
In appendix \ref{circuit::spin::boson}, for completeness, we sketch the main characteristics of the discrete model in the context of circuit-QED.   
Of course, when possible, the convergence of this model to the Ohmic (continuum) model has been verified.
\section{Two-qubit spin-boson model: static properties}
\label{two::qubits::static}
\begin{figure}[h!]
    \centering
    \includegraphics[scale=0.4]{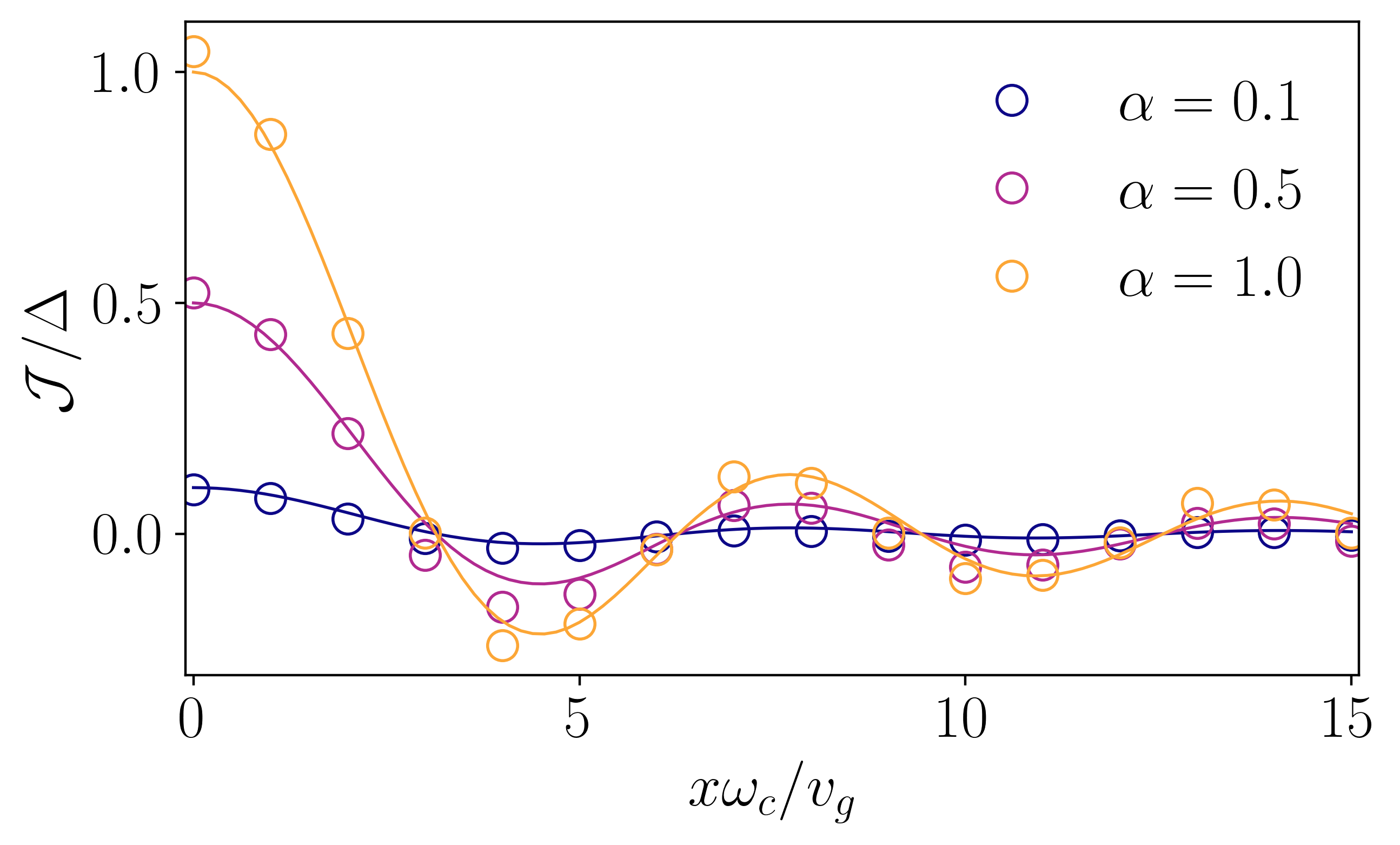}
    \caption{Ising-like coupling as a function of the normalized relative distance between the emitters, for different coupling parameters. Solid lines are plotted according to Eq. (\ref{J::sinc}) and empty dots correspond to numerical results based on the discrete Ohmic model with parameters:$\left\{N,\Delta/\omega_c\right\}$$=\left\{1001,0.2\right\}$.}
    \label{fig1::ising}
\end{figure}

The ground state of two qubits in an Ohmic environment was discussed in \cite{McCutcheon2010}. 
They used a variational ansatz equivalent to the variational polaron treatment. In this section, we review ground state properties, that will be needed when discussing the dynamics of the system. Besides, this is useful to benchmark the discrete model for numerical simulations.
For the two-qubit case, the relative distance between emitters plays an important role in the static and dynamical properties. In this situation the induced Ising interaction of the effective Hamiltonian modifies the single-qubit behavior, giving rise to collective effects. 
For $N_q=2$ the Hamiltonian (\ref{spin::hamiltonian}) reads
\begin{eqnarray}
\label{two::spin::hamiltonian}
H_{S}=\frac{\Delta_r}{2}\sum_{j=1}^{2}\sigma^{z}_{j}-\mathcal{J}\sigma^{x}_{1}\sigma^{x}_2+2\sum_{k}f_{k}(\omega_k f_k-2g_k), \nonumber\\
\end{eqnarray}
where $\mathcal{J}(x)=2\sum_k
f_{k}\left(2g_{k}-\omega_k f_k\right)\cos(kx),$ being $x=x_2-x_1$ the distance between the qubits. In this case the ground state energy is $E_{GS} = -\sqrt{\Delta_r^2+\mathcal{{J}}^2}+2\sum_{k}f_{k}\left(\omega_k f_{k}-2g_{k}\right)$, and its minimization results in
\begin{eqnarray}
\label{f_k::two::qubits}
f_k = \frac{g_k}{\omega_k}\frac{\mathcal{E}+\mathcal{J}\cos(kx)}{\mathcal{E}+\mathcal{J}\cos(kx)+\Delta_r^2/\omega_k},
\end{eqnarray}
where $\mathcal{E}=\sqrt{\Delta_r^2+\mathcal{J}^2}$.  
The effective Ising model is characterized by the ratio ${\mathcal J}/ \Delta_r$.  
In what follows, we compute both parameters numerically
by using the finite-$N$ model.  
However, it is always appealing to have analytical expressions even if approximates.
Taking the limit $\mathcal{J}\rightarrow 0$ in Eq. (\ref{f_k::two::qubits}) we recover the one-qubit relation in Eq. (\ref{f_k::one::qubit}). 
Interestingly, the induced Ising coupling can be computed in the large coupling approximation for which $f_{k}\xrightarrow{}g_{k}/\omega_k$
as $\Delta_r\rightarrow0$, which gives the result \cite{McCutcheon2010}
\begin{eqnarray}
\label{J::sinc}
\mathcal{J}&\approx&\frac{1}{\pi} \int_{0}^{\omega_c}d\omega \frac{J(\omega)}{\omega}\cos(\omega x/v_g)\nonumber\\
&=&\alpha\omega_c{\rm{sinc}}(\omega_c x/v_g),
\end{eqnarray}
where we have used the Ohmic spectral function $J(\omega)=\pi\alpha\omega$, and assumed a linear dispersion relation $\omega_k=v_g|k|$ in the continuum limit. This Ising coupling shows long-range damped oscillations as a function of the distance. This means that, depending on the distance, ferromagnetic ($\mathcal{J}>0$) or antiferromagnetic ($\mathcal{J}<0$) interaction can be induced. 
In Fig. \ref{fig1::ising} we show the behavior of the Ising-like coupling as a function of the emitters relative distance for different coupling strengths values. Continuous lines are calculated using the approximation in  Eq. (\ref{J::sinc}) and open circles indicate results obtained from numerical simulations based on the microscopic discrete model presented in the appendix (\ref{discrete::model}). We observe a very good agreement between the continuous and the discretized microscopic Ohmic model. A more accurate but cumbersome expression for the Ising coupling can also be obtained by using the exact expression for $f_k$ in Eq. (\ref{f_k::one::qubit}), which includes small correction terms to the formula given in Eq. (\ref{J::sinc}). We show this expression in appendix (\ref{Ising::long}).

Obtaining an expression for the renormalized qubit frequency $\Delta_r$ for two qubits is not that simple as in the single qubit case.
In general, 
it has to be found by numerical means. Assuming that $\mathcal{J}$ is given by the Eq. (\ref{J::sinc}), and therefore does not play any role on the energy minimization, the optimal variational parameters can be approximated by $f_{k}\approx g_{k}\mathcal{E}/(\omega_k \mathcal{E}+\Delta_{r}^2)$, which after replacement into Eq. (\ref{deltar}) gives
\begin{eqnarray}
\label{deltar::equation}
\Delta_r = \Delta \exp\left[-2 \sum_{k}\frac{g_{k}^2\mathcal{E}^2}{(\omega_{k}\mathcal{E}+\Delta_r^2)^2}\right].
\end{eqnarray}
This equation shows that $\Delta_r$ is now distance-dependent trough the Ising-like coupling. For infinite separation distance ($\mathcal{J}=0$), the renormalized frequency in the scaling limit is given by $\Delta_r\approx\Delta\left(e\Delta/\omega_c\right)^{\alpha/(1-\alpha)}$, which shows the quantum phase transition at $\alpha=1$, in agreement with the single impurity case in Eq. (\ref{deltar::onequbit}). For slightly separated qubits in the limit $\mathcal{J}/\Delta\gg 1$, we can take $\mathcal{E}\rightarrow\mathcal{J}$, obtaining that
\begin{eqnarray}
\Delta_r \approx \Delta\left(e\Delta^2/\omega_c\mathcal{J}\right)^{\alpha/(1-2\alpha)},
\end{eqnarray}
predicting the critical coupling at $\alpha=1/2$ in stark contrast with a single qubit \cite{McCutcheon2010}. It is worth mentioning that this shifted critical coupling can also be inferred by studying the properties of the dynamical Kernel in the general non-Markovian case \cite{DiazCamacho2016}.
\begin{figure}
    \centering
    \includegraphics[scale=0.4]{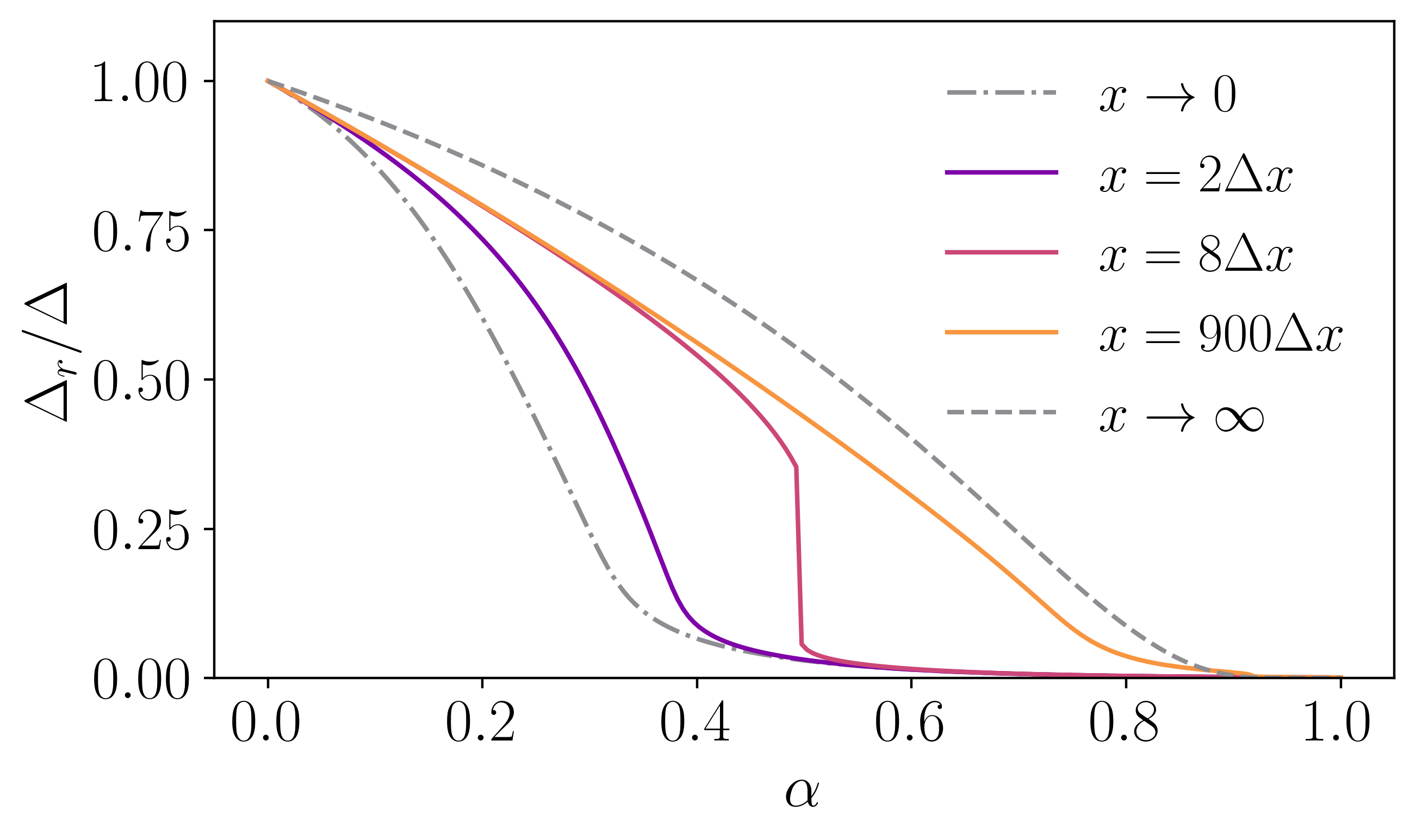}
    \caption{Renormalized frequency as function of the coupling strength for different relative distances between emitters. Oher parameters are the same as in Fig. \ref{fig1::ising}.}
    \label{fig3}
\end{figure}
In Fig. \ref{fig3} we plot the renormalized frequency varying the relative distance $x$. We show the lower and upper bounds, given by the expressions for zero and infinite separation distances, respectively. It is clear that the effective interaction between qubits modifies the localization transition point depending of their relative position. In particular, at intermediate relative positions (e.g. $x=8\Delta x$ in Fig. \ref{fig3}), $\Delta_r$ drops abruptly to zero in a discontinuous way, and the localization transition occurs for values of the coupling strength $\alpha\in\left[0.5, 1
.0\right]$.
\subsection{Ground State and Entanglement Entropy}
\begin{figure}
    \centering
    \includegraphics[scale=0.4]{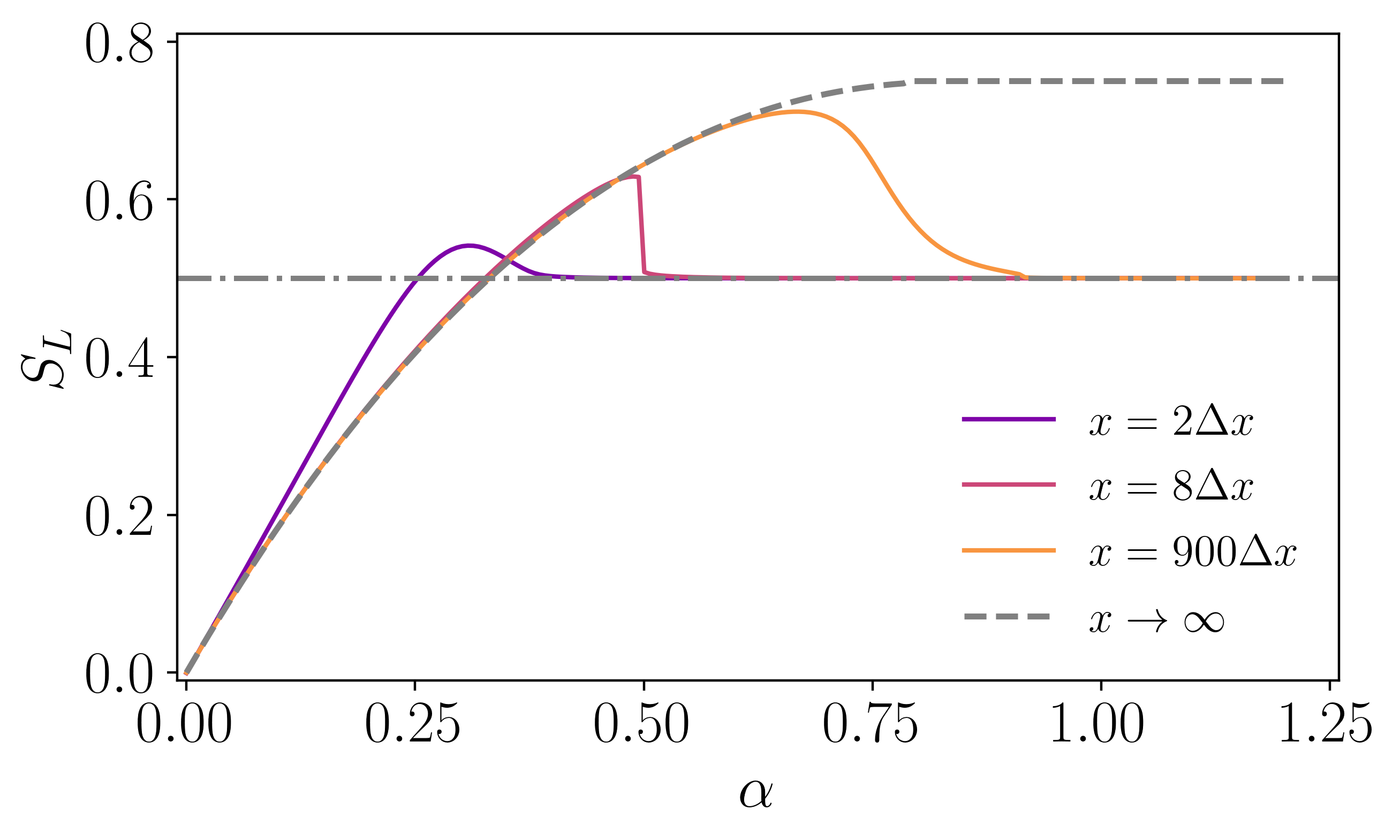}
    \caption{Entanglement entropy as a function of the coupling strength for increasing distance between the emitters. Parameters are the same as in Fig. \ref{fig3}.}
    \label{fig4}
\end{figure}

The ground state of the effective two-spin Hamiltonian (\ref{two::spin::hamiltonian}) can be found by straightforward diagonalization. It reads
\begin{eqnarray}
\ket{\rm GS}_{S}=\cos\theta\ket{gg}+\sin\theta\ket{ee},
\end{eqnarray}
where 
\begin{eqnarray}
\cos\theta=\frac{\Delta_r+\mathcal{E}}{\sqrt{(\Delta_r+\mathcal{E})^2+\mathcal{J}^2}},
\end{eqnarray}
and 
\begin{eqnarray}
\sin\theta=\frac{\mathcal{J}}{\sqrt{(\Delta_r+\mathcal{E})^2+\mathcal{J}^2}}.
\end{eqnarray}
The ground state of the complete system of two-qubits plus the bosonic bath is then obtained by applying the polaron transformation to the product of the ground state $\ket{\rm GS}_{S}$ and the multimode vacuum state of the bosonic environment according to the ansatz (\ref{ground::state::ansatz})
\begin{eqnarray}
\ket{\Psi_{\rm GS}}&=& U_{P}\ket{\rm GS}_S \ket{\mathbf{0}}\nonumber\\
&=&\left(\cos\theta\sigma^{x}_1+\sin\theta\sigma^{x}_2\right)U_{P}\ket{e}\ket{g}\ket{\mathbf{0}}.
\end{eqnarray}
After applying the transformation we get the following entangled state in terms of multimode coherent states of the photonic bath
\begin{eqnarray}
\ket{\Psi_{\rm GS}}&=&\frac{1}{2}\left(\sin\theta+\cos\theta\right)\ket{--}\ket{\chi_{--}}  \nonumber\\ &+&\frac{1}{2}\left(\sin\theta-\cos\theta\right)\ket{-+}\ket{\chi_{-+}}\nonumber\\
&+& \frac{1}{2}\left(\sin\theta-\cos\theta\right)\ket{+-}\ket{\chi_{+-}}\nonumber\\
&+& \frac{1}{2}\left(\sin\theta+\cos\theta\right)\ket{++}\ket{\chi_{++}},
\end{eqnarray}
where we chose the eigenstates of $\sigma_{j}^x$ for representing the state of the qubits, i.e.,  $\ket{\pm}_{j}=(\ket{e}_{j}\pm\ket{g}_{j})/\sqrt{2}$, and the multimode coherent states are given by
\begin{eqnarray}
\ket{\chi_{\pm\pm}}&=&\exp\left[-\imag\sum_{k}f_{k}^2\sin kx\right]\bigotimes_{k}\ket{\pm (f_k+ f_k e^{\imag k x})},\quad\quad\\
\ket{\chi_{\pm\mp}}&=& \exp\left[\imag\sum_{k}f_{k}^2\sin kx\right]
\bigotimes_{k}\ket{\pm f_k \mp f_k e^{\imag k x}},
\end{eqnarray}
with $\ket{\alpha_k}=D(\alpha_k)\ket{\mathbf{0}}$, being $D(\alpha_k)$ the displacement operator.  
Notice that in the deep-strong coupling (localized regime) when $\Delta_r\rightarrow0$, the ground state becomes
$\ket{\Psi_{\rm GS}}=\left(\ket{--}\ket{\chi_{--}}+\ket{++}\ket{\chi_{++}}\right)/\sqrt{2}$, which can be recognized as a Schr\"{o}dinger cat state of light and matter analogous to the one-qubit SBM \cite{florens2017}. Having obtained the ground state, one can also compute relevant physical observables. For instance,  the qubit magnetization and the entanglement entropy. In the first case we find that 
\begin{eqnarray}
\ave{\sigma^{z}_j}_{\rm GS}=\bra{\Psi_{\rm GS}}\sigma^{z}_j\ket{\Psi_{\rm GS}}
=-\frac{\Delta_r}{\Delta}\cos2\theta\cos\left[\phi(x)\right],\quad
\end{eqnarray}
where we have introduced the auxiliary function $\phi(x)=4\sum_{k}f^{2}_{k}\sin k x$. It is also interesting to compute the linear entanglement entropy, defined as $S_{L}=1-\rm{Tr}\varrho_{\rm q}^2$, where the reduce density matrix of the two-qubit system can be obtained by tracing out the bath degrees of freedom, i.e., $\varrho_{\rm q}=\rm{Tr}_{\rm env}\varrho_{\rm GS}=\rm{Tr}_{\rm env}\proj{\Psi_{\rm GS}}$. This measure quantifies the degree of mixedness of the two-qubit ground state with the bosonic bath. 
Remarkably, a rather general expression of the linear entropy for arbitrary emitter separations can be obtained, but the expression is too lengthy to be shown here (see appendix \ref{entropy::large}). However, we show the limiting formula for infinite distance
\begin{eqnarray}
\label{infinite::entropy}
S_{L}=\frac{1}{4}\left[3-
2\left(\frac{\Delta_r}{\Delta}\right)^2-\left(\frac{\Delta_r}{\Delta}\right)^4\right].
\end{eqnarray}
It is well known that entanglement entropy reflects the appearance of a quantum phase transition, so we expect to observe signals of the localization transition in the entanglement as the coupling is increased, in a similar fashion as observed for $\Delta_r$ in Fig. \ref{fig3}. Quantum phase transitions manifest as a nonanalyticity in the entanglement contained in the total state of the system and its environment \cite{Brandes2005}. We confirm this by plotting the entanglement entropy as a function of the coupling in Fig. \ref{fig4} for increasing relative distance. The dashed line in Fig. \ref{fig4} is given by Eq. (\ref{infinite::entropy}), indicating a smooth behavior of the entropy with no discontinuities for large couplings. This is expected from the one-qubit results for the renormalized frequency discussed in section \ref{theoretical::model} for which $\alpha_c\sim 1$. The asymptotic entropy in this case is a complete mixed state of the qubits with $S_{L}\rightarrow 3/4$, which can be seen directly from Eq. (\ref{infinite::entropy}) in the limit $\Delta_r\rightarrow 0$. For finite separation we have a richer behavior, as sudden decay of the entropy occurs at specific couplings, reflecting the influence of the effective qubit-qubit interaction on the localization transition. Notice also that in this case the steady state entropy as a function of the coupling is $S_{L}=1/2$. This is because the Ising interaction dominates over the magnetic field produced by $\Delta_r$ in the two-spin effective model, and the ground state becomes degenerate with reduced density matrix $\varrho_{\rm q}=\mathds{1}/2$. This is also clear from the general expression for the linear entropy in Eq. (\ref{entropy::general}) in the case of $\sin\theta=\cos\theta=1/\sqrt{2}$.

\section{Dynamics}
\label{sect:dyn}
Not only the ground state but the first excited states are well captured by the polaron ansatz \cite{Bera2014, DiazCamacho2016, sanchez2019,JRR2020}.
It is convenient to work in the \emph{polaron picture} (\ref{spin::boson}): $H_{P}=U^{\dagger}_{P}HU_{P}\approx H_{\rm eff}$,
which can be approximated by \cite{DiazCamacho2016}
\begin{eqnarray}
    H_{\rm eff}&=&\frac{\Delta_r}{2}\sum_{j=1}^{2}\sigma^{z}_{j}-2\Delta_r\sum_{j=1}^2\sigma^{z}_j\sum_{kk^\prime}f_{k}f_{k^\prime}e^{i(k-k^\prime)x_{j}}a^{\dagger}_{k}a_{k^\prime}\nonumber\\
    &+&
    \sum_{j=1}^{2}\sum_{k}\left[2\Delta_r f^{(0)}_{k}+\epsilon\left(\Delta_r-\omega_k\right)\right]\left(\sigma^{-}_{j}a^{\dagger}_{k}e^{\imag k x_{j}}+\rm{h.c.}\right) \nonumber\\
    &+& \sum_{k}\omega_k a^{\dagger}_{k}a_{k}-\mathcal{J}\sigma^{x}_{1}\sigma^{x}_2+2\sum_{k}f_{k}(\omega_k f_k-2g_k). \nonumber\\
\end{eqnarray}
Here $f^{(0)}_{k}$ is the variational parameter of a single qubit given by Eq. (\ref{f_k::one::qubit}), i.e., when the ferromagnetic interaction is negligible ($\mathcal{J}=0$), and $\epsilon$ is a small correction. 
The key fact here is that $H_{\rm eff}$  {\it preserves the excitation number}.  Therefore, this effective Hamiltonian can be now treated using standard RWA methods. In the following we will study the generated dynamics in the polaron frame in order to extract relevant information of the system in the lab picture. 
\\

As the Hamiltonian $H_{\rm eff}$ preserves the number of total excitations, we can project the dynamics in the  single excitation sector  spanned by the basis: $\{\ket{e}\ket{g}\ket{\mathbf{0}},\ket{g}\ket{e}\ket{\mathbf{0}},\ket{\rm GS}_{S}\ket{1_{k}}\}$. This means that photonic excitations are created from the polaron ground sate of the two-spin system.
The state of the system at any time is postulated via the following dynamical polaron \emph{ansatz} \cite{DiazCamacho2016},
\begin{eqnarray}
\label{dynamical::polaron}
    \ket{\Psi(t)}_{P}&=&\left(c_{1}\ket{e}\ket{g}+c_{2}\ket{g}\ket{e}+\sum_{k}\psi_{k}a^{\dagger}_{k}\ket{\rm GS}_{S}\right)\ket{\mathbf{0}},\nonumber\\
\end{eqnarray}
where the time-dependent probability amplitudes require to fulfill the normalization condition $\sum_{i}|c_{i}|^2+\sum_{k}|\psi_{k}|^2=1$.
These coefficients satisfy the coupled set of equations,
\begin{eqnarray}
\label{c1::equation}
\imag \dot c_{1}&=&-\mathcal{J}c_{2}+\sum_k G_ke^{-\imag kx_{1}}\psi_k\cos\theta, \\
\label{c2::equation}
\imag\dot c_{2}&=&-\mathcal{J}c_{1}+\sum_k G_ke^{-\imag kx_{2}}\psi_k\cos\theta, \\
\label{psik::field}
\imag\dot\psi_k&=&-(\tilde\Delta_r-\omega_k)\psi_k+G_k\sum_{j}e^{\imag kx_{j}}c_{j}\cos\theta \nonumber\\
&+&2\Delta_r f_k\sum_{j,k^\prime}f_{k^{\prime}}e^{\imag(k-k^{\prime})x_{j}}\psi_{k^{\prime}}\cos2\theta.
\end{eqnarray}
Here we have defined the functions $G_k=2\Delta_r f^{(0)}_{k}+\epsilon(\Delta_r-\omega_k)$, and $\tilde\Delta_r=\Delta_r\cos2\theta+\mathcal{J}\sin2\theta.$
Using the discrete model, this set of equations are integrated numerically and, then, transformed back to the original lab picture.   
\subsection{Dynamics in the polaron frame}
For future convenience and understanding, it is useful to start analyzing the dynamics in the polaron picture.
The reason is twofold.  It is easier, so some explicit formulae can be obtained, and it will be convenient to understand the dynamics in the lab frame.
Besides, if the coupling is small enough, both polaron and polaron frame are equivalent.
Moving to the rotating frame for the photon  amplitudes, $\tilde\psi_k=e^{-\imag(\tilde\Delta_r-\omega_k)t}\psi_k$, and performing the integration, i.e., tracing out the photonic modes, and setting $\tilde\psi_k(0)=0$, we arrive to 
\begin{widetext}
\begin{eqnarray}
\label{delay::general}
\imag\dot c_{i}&=&-\mathcal{J}c_{j}-\frac{2\imag\Delta_r^2\cos^2\theta}{\pi}\int_{0}^{\omega_c} d\omega\frac{J(\omega)}{(\omega+\Delta_r)^2}\int_{0}^{t}d\tau
e^{\imag(\tilde\Delta_r-\omega)\tau}c_{i}(t-\tau)\nonumber\\
&-&\frac{2\imag\Delta_r^2\cos^2\theta}{\pi}e^{\imag\tilde\Delta_r x/v_g}\int_{0}^{\omega_c} d\omega\frac{J(\omega)}{(\omega+\Delta_r)^2}\int_{0}^{t-x/v_g}
d\tau^{\prime}e^{\imag(\tilde\Delta_r-\omega)\tau^{\prime}}c_{j}(t-x/v_g-\tau^{\prime}),
\end{eqnarray}
\end{widetext}
where it was assumed that $|G_{k}|^2\approx4\Delta_r^2{f^{(0)}_{k}}^{2}$ for small $\epsilon$. 
Here,  the presence of non-Markovian processes is evident. The second and third contributions in the r.h.s both depend on the spectral function $J(\omega)$ of the waveguide (environment). So, there is an unavoidable source of non-Markovianity as consequence of the memory of the environment, which cannot be considered as flat (Markovian) in general \cite{breuer}.
%
%
On the other hand, the third term contains an additional source of non-Markovianity  due to the finite distance effect between emitters, which also enters in the non-local parts of the Kernel and therefore invalidates the Markov approximation. These two non-Markovian time scales are different in general, and need to be distinguished in the dynamical evolution. Unfortunately, this is not so simple and we have to get rid of one of such memory time scales in order to study the behavior of Eq. (\ref{delay::general}).  
At this point we follow the lines of the Weisskopf-Wigner theory of spontaneous emission \cite{Weisskopf1930}, and neglect the memory effects of the environment in order to evaluate the contribution of the last two terms in Eq. (\ref{delay::general}). We can then make the replacement $c_{i}(t-\tau)\approx c_{i}(t)$ and extend the time integral to infinity. By using the identity \cite{CarmichaelBook}
\begin{eqnarray}
\lim_{t\to\infty}\int_{0}^{t}d\tau e^{i(\tilde\Delta_r-\omega)\tau}=\pi\delta(\omega-\tilde\Delta_r)+i\mathcal{P}\left(\frac{1}{\tilde\Delta_r-\omega}\right), \nonumber\\
\end{eqnarray}
we arrive to the coupled delay-differential equations
\begin{eqnarray}
\label{delay::equations}
\dot c_{i}(t)&=&\imag\mathcal{J}c_{j}(t)
-\frac{\gamma}{2}c_{i}(t)\nonumber\\
&-&\frac{\gamma}{2}e^{\imag\tilde\Delta_r x/v_g}c_{j}(t-x/v_g)\Theta(t-x/v_g),
\end{eqnarray}
with the spontaneous emission rate defined by
\begin{eqnarray}
\gamma = \frac{4\Delta_r^2J(\tilde\Delta_r)}{\left(\Delta_r+\tilde\Delta_r\right)^2}\cos^{2}\theta,
\end{eqnarray}
and the function $\Theta(\tau)$ being the Heaviside step function.
Obviously, once the approximation $c_{i}(t-\tau)\approx c_{i}(t)$ is made, the only source of non-Markovianity is due to the finite distance of the emitters, which induces backaction of the field, leading to a radiation feedback phenomenon in the system. 
As a consequence, these equations resemble the delay dynamics obtained for two distant emitters coupled to a waveguide with flat spectral density of the field modes \cite{Kanu19}. However, in this case a coherent coupling term between the emitters appears as a consequence of the Ising coupling $\mathcal{J}$ induced by the polaron transformation.
As this coherent interaction depends on the distance between emitters, see Eq. \eqref{J::sinc} and Fig. \ref{fig1::ising},  it contributes to the dynamics at small separation distances. Notice also that the spontaneous emission rate and the relative phase gained by the retarded radiation gets renormalized as they are given in terms of the polaron qubit frequency $\Delta_r$. 
For infinitely distant emitters, the spontaneous emission rate reduces to the result of a single emitter with $\gamma=J(\Delta_r)=\pi\alpha\Delta_r$ \cite{DiazCamacho2016,Zueco2019}.

It is worth mentioning that in deriving the delay differential equations (\ref{delay::equations}), we have neglected the Lamb-shift correction to the qubit frequency which is formally given by
\begin{eqnarray}
\delta_L=\frac{2}{\pi}\Delta_r^2\mathcal{P}\left(\int_0^{\omega_c}d\omega \frac{J(\omega)}{\left(\omega+\Delta_r\right)^2\left(\tilde\Delta_r-\omega\right)}\right).
\end{eqnarray}
A detailed analysis of the consequences of collective spontaneous emission and Lamb-shift is given in \cite{DiazCamacho2015}.  
The magnetization dynamics (or population inversion) in the polaron image can be expressed in terms of the qubit probability amplitudes,
\begin{eqnarray}
    \ave{\sigma^{z}_{i}(t)}_{P}=\cos^2\theta\left(2|c_i(t)|^2-1\right)-\sin^2\theta\left(2|c_j(t)|^2-1\right). \nonumber\\
\end{eqnarray}
Let us now analyze the emergence of non-Markovian dynamics due to the back-action induced by the delay term in Eqs.(\ref{delay::equations}) for different initial states of the emitters. Naturally, we expect these results to agree with those obtained within the RWA with a flat environment for very small couplings \cite{Kanu19}. 
We first explore the case of large but not infinite separation distance between the emitters for which $\mathcal{J}\approx0$. The steady state solutions in the polaron picture can be computed from the standard final value theorem \cite{Chen2007}, by taking the Laplace transform in Eqs.(\ref{delay::equations}). We first take initial symmetric (antisymmetric) states in the polaron {\it ansatz}, i.e.,  $c_{1}(0)=c_{2}(0)=1/\sqrt{2}$ or $c_{1}(0)=-c_{2}(0)=1/\sqrt{2}$. For these initial conditions the Laplace transformed amplitudes read
\begin{eqnarray}
\tilde c_{\pm}(s)=\frac{1}{\sqrt{2}\left[s+\gamma/2\pm\gamma e^{(\imag\Delta_r-s)x/v_{g}}/2\right]},
\end{eqnarray}
where $\tilde c_{+}=\tilde c_{1}=\tilde c_{2}$, and $\tilde c_{-}=\tilde c_{1}=-\tilde c_{2}$. For a renormalized qubit frequency, and separation distance satisfying the condition $\Delta_r x/v_g=2n\pi$, being $n\in\mathbb{Z}$, 
the symmetric initial state will decay to the polaron ground state, i.e., $\lim_{t\to\infty}c_{+}(t)=0$. However, if the initial state is antisymmetric we get a bound state with finite excitation probability amplitude given by $\lim_{t\to\infty}c_{-}(t)=\left(1+\gamma x/2v_g\right)^{-1}/\sqrt{2}$ \cite{Kanu19}.
This gives the steady state magnetization for the antisymmetric state $\ave{\sigma^z_-}_{P}=\left(1+\gamma x/2v_g\right)^{-2}-1$, which is different from $-1$ for finite-distance emitters (see Fig. \ref{fig:sigmaz:supersub}(a)). This bound state in the continuum (BIC) is the result of a trapped stationary excitation between the two emitters acting as an effective cavity formed by two perfectly reflective mirrors \cite{Tuffarelli2013,Calajo2019}.

\subsection{Lab frame}

The previous analysis will help us discuss the actual dynamics.
They can be obtained by back-transforming with the polaron unitary. It turns out this can be done exactly, as we have restricted our analysis to the single-excitation polaron manifold. In particular, the magnetization in the lab picture reads
\begin{eqnarray}
\ave{\sigma_{i}^{z}(t)}&=&{}_{P}\bra{\Psi(t)}U^{\dagger}_P\sigma_{i}^z U_{P}\ket{\Psi(t)}_{P} \nonumber\\
&=&\frac{\Delta_r}{\Delta}\left[\ave{\sigma^{z}_{i}}_{P}+4\cos\theta\Re\left\{c_{i}\sum_k f_k\psi^{*}_k\right\}\right. \nonumber\\
    &-&\left.  4\sin\theta\Re\left\{c_{j}\sum_k f_k\psi^{*}_k\right\}\right.\nonumber\\
    &+&\left. 4\cos2\theta\sum_{kk^{\prime}}f_{k}f_{k^{\prime}}\psi_k\psi^{*}_{k^{\prime}}\right],
    \label{szlab}
\end{eqnarray}
which shows the dynamical contributions due to each emitter and field correlations. 

In \eqref{szlab} we see that the differences between both frames are ${\mathcal O} (f_k)$, therefore in the weak coupling both frames give the same results, as expected by construction.
In the same region, one can perform the RWA approximation already in the original spin boson model \eqref{spin::boson}.
This is illustrated in Fig. \ref{fig:sigmaz:supersub}(a) for which we recover RWA results. 

As we increase the system-bath coupling $\alpha$, non-Markovian dynamics emerges as consequence of the memory of the bath and also from interference caused by retardation effects \cite{Kanu19}. At this point the collective two-qubit dynamics is no longer captured by the effective delay dynamics dictated by Eqs. (\ref{delay::equations}). We show these results in Fig. \ref{fig:sigmaz:supersub}(b), (c) and (d) for a fixed separation distance between the emitters. 
We also expect that for very strong couplings, the  localization transition takes place ($\Delta_r\rightarrow0$), and the two-qubit system freezes in a state of zero magnetization. In this deep-strong coupling regime retardation effects are no longer present in the magnetization. 

\subsubsection{The Fermi Problem in the USC regime}

\begin{figure}
    \centering
    \includegraphics[width = \columnwidth]{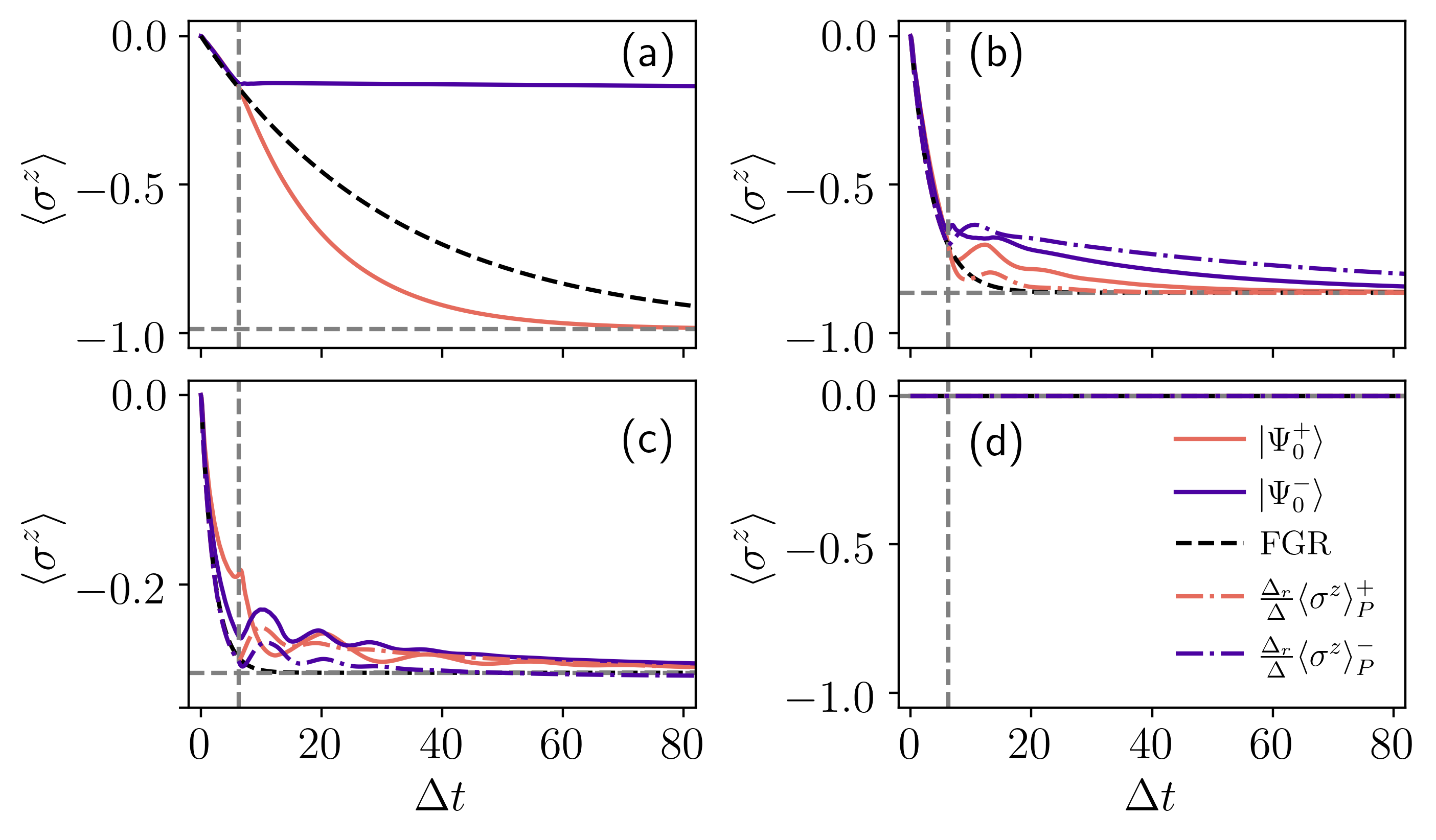}
    \caption{Dynamics of the magnetization $\ave{\sigma^z}$ for initial symmetric $\ket{\Psi_0^+}$ and antisymmetric $\ket{\Psi_0^-}$ states. Dash-dotted lines indicate the evolution in the polaron picture. To enable comparison with the true dynamics, the polaron lines are renormalized by $\Delta_r / \Delta$. From (a) to (d) the increasing couplings are $\alpha = 0.01$ , $\alpha = 0.1$, $\alpha = 0.5$ and $\alpha = 1.0$. Other parameters are $L=40\pi$, $N=1001$, $x=2\pi v_{g}/\Delta$.}
    \label{fig:sigmaz:supersub}
\end{figure}

\begin{figure}
    \centering
    \includegraphics[width = \columnwidth]{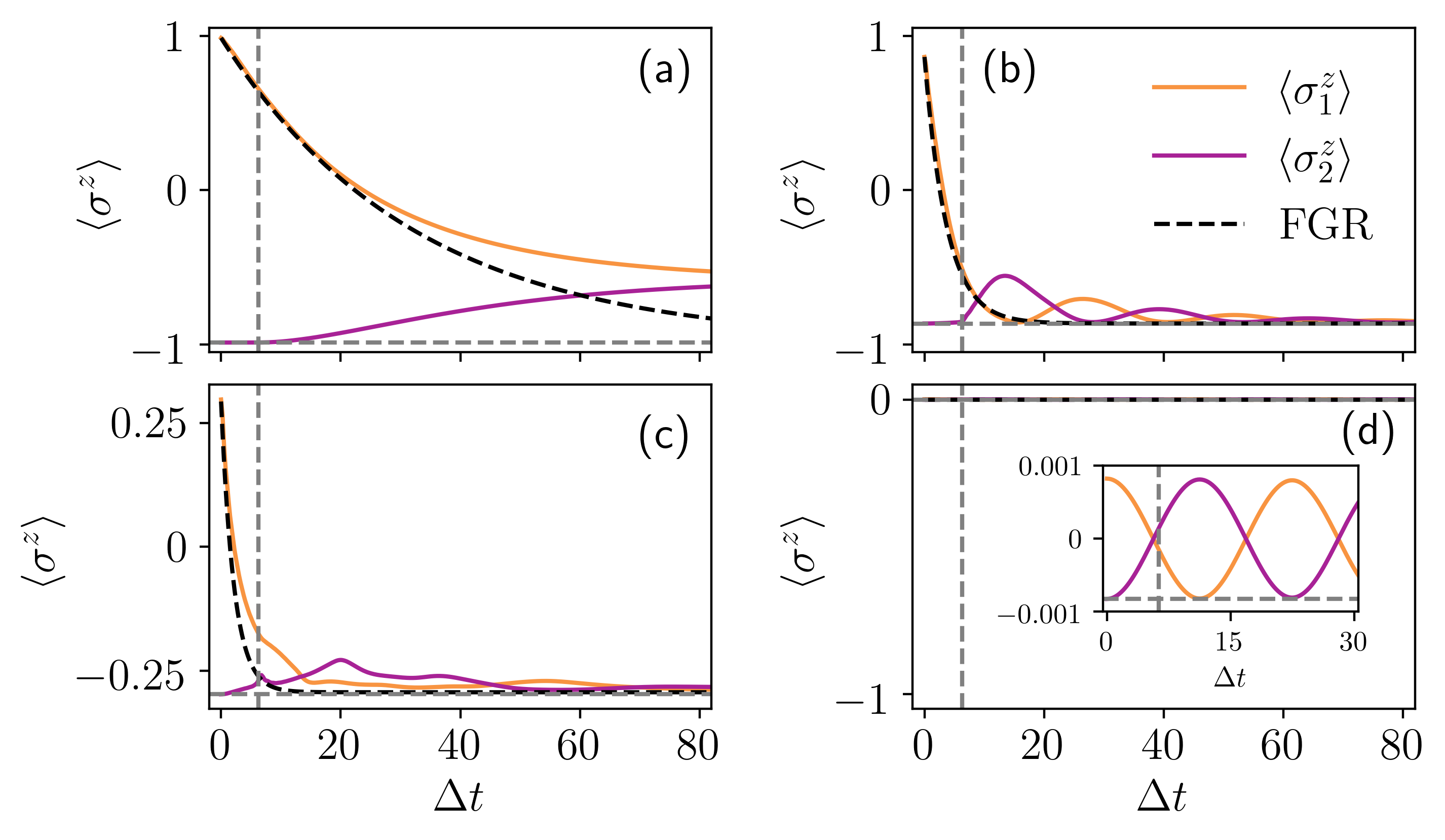}
    \caption{Dynamics of $\ave{\sigma_{1,2}^z}$ starting from the initial state $\ket{eg}$. From (a) to (d) the parameters are the same as in Fig. \ref{fig:sigmaz:supersub}}
    \label{fig::eg::state}
\end{figure}

Another interesting fact is the apparent causality violation that can be observed in Fig. \ref{fig:sigmaz:supersub}(c). For moderate couplings ($\alpha=0.01, 0.1$), where RWA still applies, the dynamics of symmetric  and antisymmetric initial states are exactly the same before the retardation time, i.e., independent exponential decay of both emitters at the same rate (Fermi golden rule) occurs. As we increase the coupling we can see from Fig. \ref{fig:sigmaz:supersub}(c) that this is no longer the case. Symmetric and antisymmetric initial states start to have different dynamics long before they reach the vertical line indicating the light-cone overlap between the two emitters. This apparent paradox was first studied by Fermi in 1932 and it is know as the \emph{Fermi problem} \cite{Fermi1932}. Fortunately this paradox can be explained in terms of correlations  between space-like distant events arising in the ultrastrong coupling regime and that are absent in usual RWA models \cite{Sabin2011}. The Fermi paradox can be formulated explicitly as follows: in a system of two distant emitters, where initially one of them is in the excited sate and the other is in its ground state, the following interesting question arises: is it possible to excite the second emitter trough the spontaneous emission of the first one at a time $t<x/v_g$? It turns out that this is not possible, and there is no causality violation in this problem, as was pointed out by Fermi \cite{Fermi1932}, and better justified latter by several authors \cite{Hegerfeldt1994,Buchholz1994,Power1997}. However as was shown by Sabin {\it et. al.} \cite{Sabin2011}, if non-local correlations are shared initially by the two emitters, the probability of finding the second emitter in the excited state and the first one in the ground state can be different from zero even at times $t<x/v_g$. We will show here that the existence of these correlations explains the different dynamics experienced by symmetric and antisymmetric initial states at times before the retardation time observed in Fig. \ref{fig:sigmaz:supersub}. We now move for convenience to the Heisenberg picture, and write the equation of motion for the magnetization using the original lab Hamiltonian (\ref{spin::boson})
\begin{eqnarray}
\sigma^{z}_{j}(t)=\sigma^{z}_{j}(0)+2\int_{0}^{t}d\tau \sigma^{y}_{j}(\tau)V(x_j,\tau), 
\end{eqnarray}
where we have defined
\begin{eqnarray}
V(x_j,t) &=& V_0(x_j,t)+ V_1(x_j,t)+V_{2}(x_j,t)
\end{eqnarray}
with the field operator
\begin{eqnarray}
V_0(x_j,t)=\sum_{k}g_{k}\left(a_{k}e^{-\imag k x_{j}} e^{-\imag\omega_k t}+\rm {h.c.}\right),
\end{eqnarray}
and the qubit operator
\begin{eqnarray}
V_{i}(x_j,t)=-\sum_{k}\int_{0}^t d\tau |g_{k}|^2\sigma^{x}_{i}(\tau)e^{\imag k (x_{j}-x_{i})}e^{-\imag\omega_{k}(t-\tau)}+\rm{h.c.}\nonumber\\
\end{eqnarray}
The expectation value of the qubit magnetization is then the sum of different contributions
\begin{eqnarray}
\label{sigmaz::heisenberg}
\ave{\sigma^{z}_{j}(t)}&=& \ave{\sigma^{z}_{j}(0)}+2\int_0^t d\tau\ave{\sigma^{y}_{j}(\tau)V_{0}(x_j,\tau)}\nonumber\\
&+& 2\int_{0}^{t}d\tau\ave{\sigma^{y}_{j}(\tau)V_{1}(x_{j},\tau)}\nonumber\\
&+&2\int_{0}^{t}d\tau\ave{\sigma^{y}_{j}(\tau)V_{2}(x_{j},\tau)},\nonumber\\
\end{eqnarray}
where the last two correlation terms suggest some influence or back-action of the respective distant emitter. However, it can be shown that both contributions are independent of such distant emitter \cite{Sabin2011}. To make this clear let us focus in the first emitter magnetization, i.e., $\ave{\sigma^{z}_{1}(t)}$. In this case one can show that the correlation
$\ave{\sigma^{y}_{1}(t)V_{2}(x_{1},t)}$$\propto$$\frac{d}{dx}\ave{\sigma^{y}_{1}(t)\sigma^{x}_{2}(t-x/v_g)}\Theta(t-x/v_g)$,
showing that this function is zero for times $t<x/v_g.$
An analogous calculation for the auto-correlation function of the first emitter shows that it is in fact independent of the second emitter, so the only contribution to the qubit magnetization is due to the second term in Eq. (\ref{sigmaz::heisenberg}). We can examine this contribution for initial symmetric or antisymmetric states of the emitters, which are indeed initial entangled states in the polaron picture. Then
\begin{eqnarray}
\ave{\sigma^{y}_{1}(\tau)V_{0}(x_1,\tau)}=\bra{\Psi^{\pm}_0}\sigma^{y}_{1}(\tau)V_{0}(x_1,\tau)\ket{\Psi^{\pm}_0},
\end{eqnarray}
where $\ket{\Psi^{\pm}_0}=\frac{1}{\sqrt{2}}\left(\sigma^{x}_{1}\pm\sigma^{x}_{2}\right)U_{P}\ket{g}\ket{g}\ket{\bf 0}.$ An explicit calculation of these qubit-field correlations is complicated as consequence of the non-trivial action of the polaron transform, but one can see that, in general, they are different from zero for all times and have different value for symmetric and antisymmetric initial states. This is in strong contrast with the RWA case, where these correlations are totally absent and this contribution is exactly zero, as the polaron transform does not enter in the initial state. This analysis explains the different dynamics experienced by both initial states when the system enters in the ultrastrong coupling regime as shown in Fig. \ref{fig:sigmaz:supersub}(c).  It is also worth mentioning that although we were interested in the difference in time evolution of symmetric and antisymmetric initial states, the above analysis is in fact independent of the initial state of the two-qubit system. These correlations emerge as the coupling of the two distant emitters to the Ohmic waveguide increases, breaking down the usual RWA. We have additionally explored the dynamics for the initial state $\ket{e}\ket{g}\ket{\bf 0}$ in the polaron image. This is shown in Fig. \ref{fig::eg::state} where the magnetization of each emitter is plotted for increasing coupling and for the same parameters used in Fig. \ref{fig:sigmaz:supersub}. For weak coupling we observe exponential decay (Fermi golden rule) of the excited emitter, and the subsequent excitation of the non initially excited emitter at time $t=x/v_g$. For $\alpha=0.1$ (Fig. \ref{fig::eg::state}(b)) the spontaneous decay rate gets renormalized and we are able to see the damped coherent exchange of the excitation between the two emitters until it reaches the expected equilibrium magnetization given by $-\Delta_r/\Delta.$ Excitation of the second emitter before retardation time is shown in Fig. \ref{fig::eg::state}(c) for $\alpha=0.5$ with notorious effects on the dynamics at $x/v_g$ and the expected localization transition for $\alpha=1.0$ in Fig. \ref{fig::eg::state}(d), where localized and small amplitude ($2\Delta_r/\Delta$) Rabi oscillations between the emitters are observed due to the coherent Ising interaction moving the excitation between the two emitters (see inset in Fig. \ref{fig::eg::state}(d)).
\subsubsection{Time-dependent decay rate}
\begin{figure}
    \centering
    \includegraphics[scale=0.5]{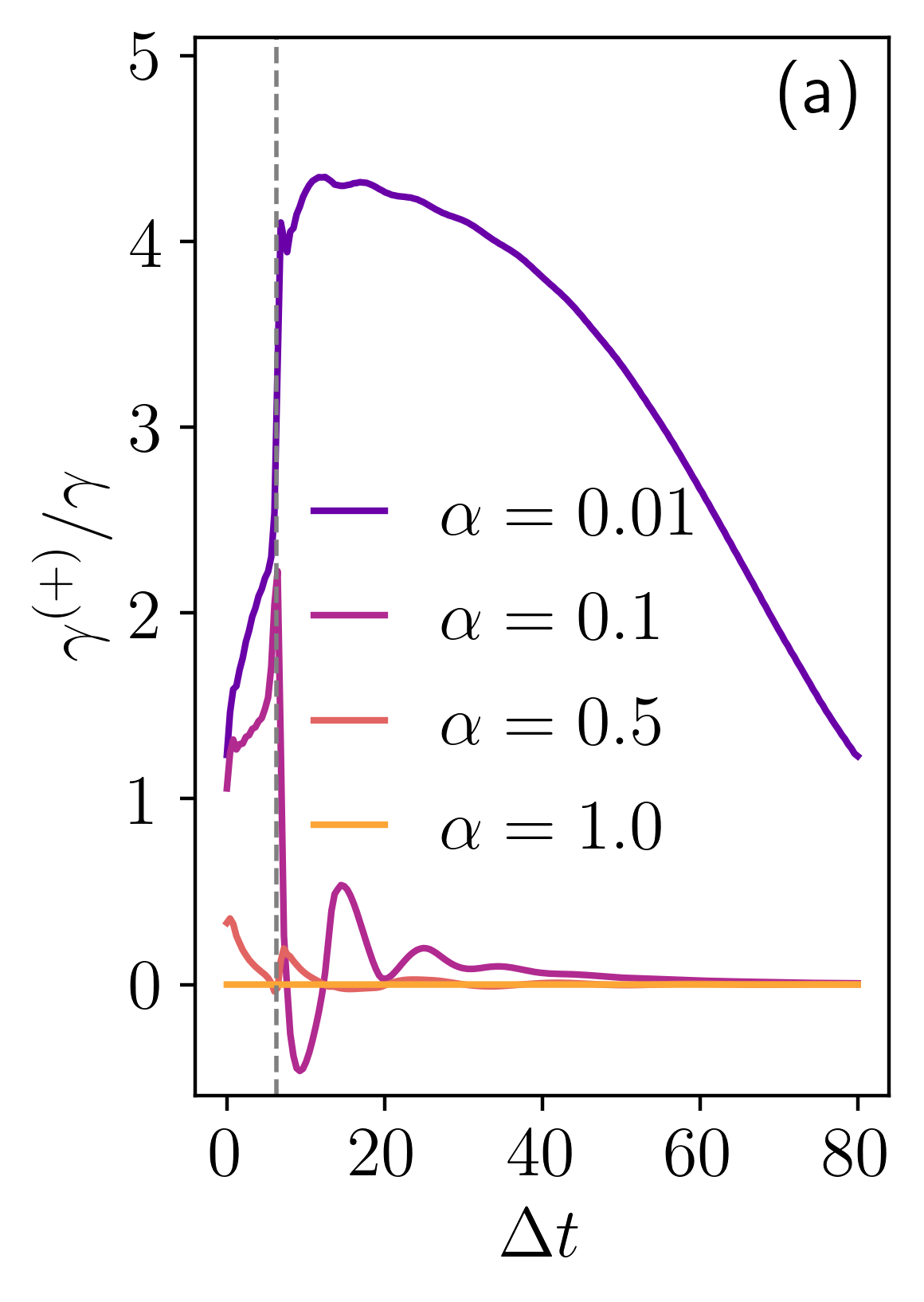}
    \includegraphics[scale=0.5]{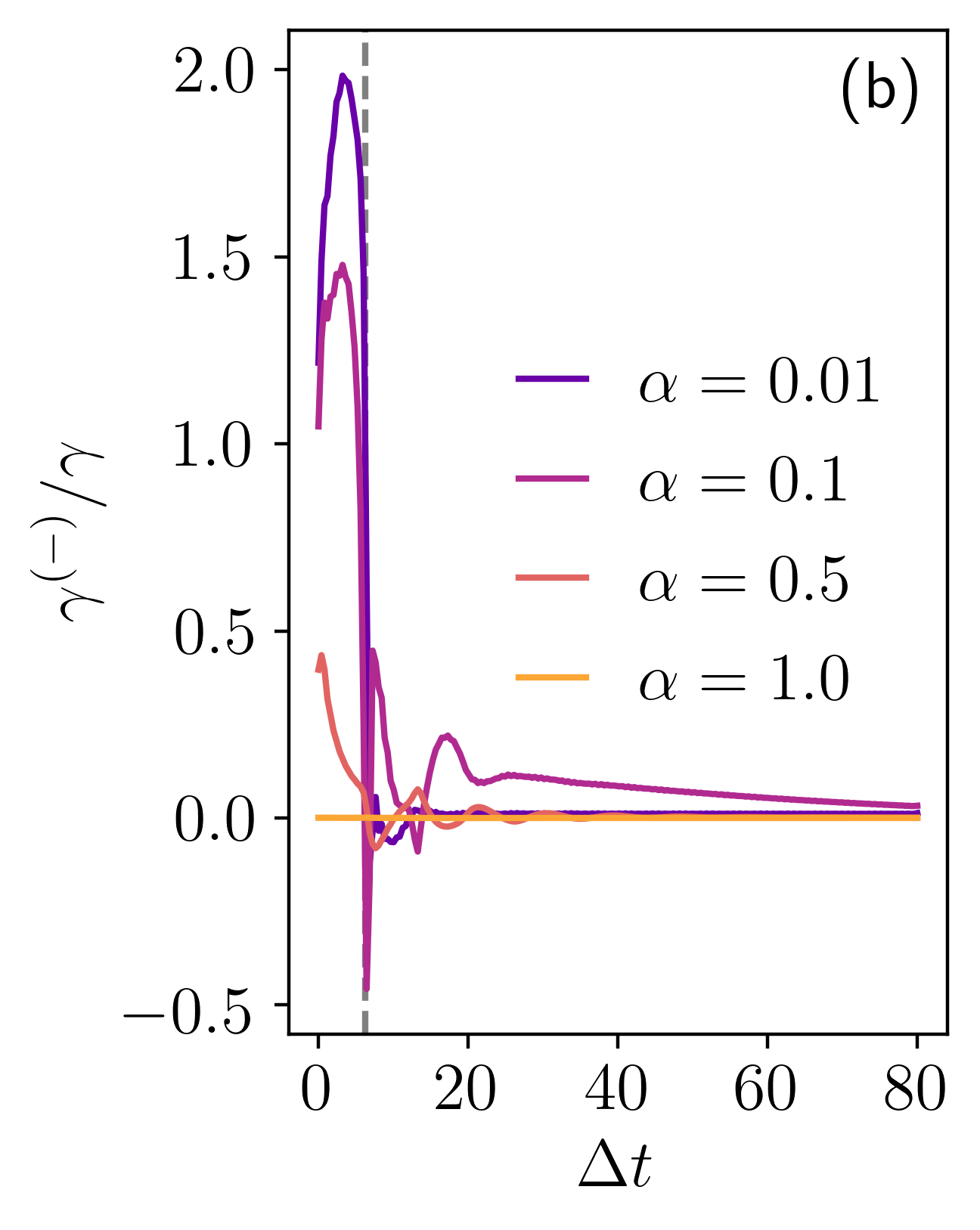}
    \caption{Dynamics of the instantaneous decay for increasing coupling $\alpha$. (a) Symmetric initial state $\ket{\Psi^{+}_0}$, and (b) antisymmetric initial state $\ket{\Psi^{-}_0}$. Other parameters are the same as in Fig. \ref{fig:sigmaz:supersub}.}
    \label{gamma::tiempo}
\end{figure}

\begin{figure}
    \centering
    \includegraphics[scale=0.5]{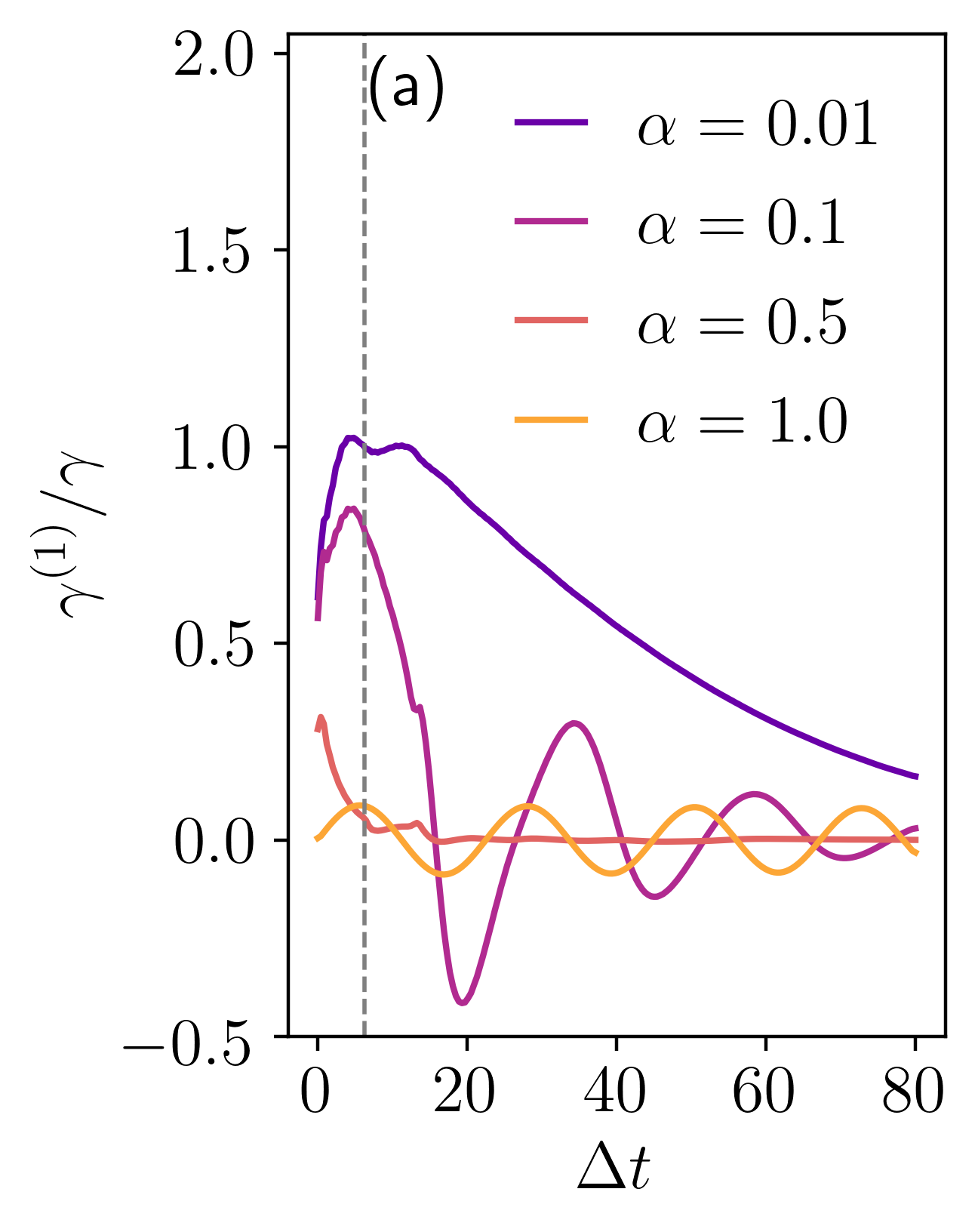}
    \includegraphics[scale=0.5]{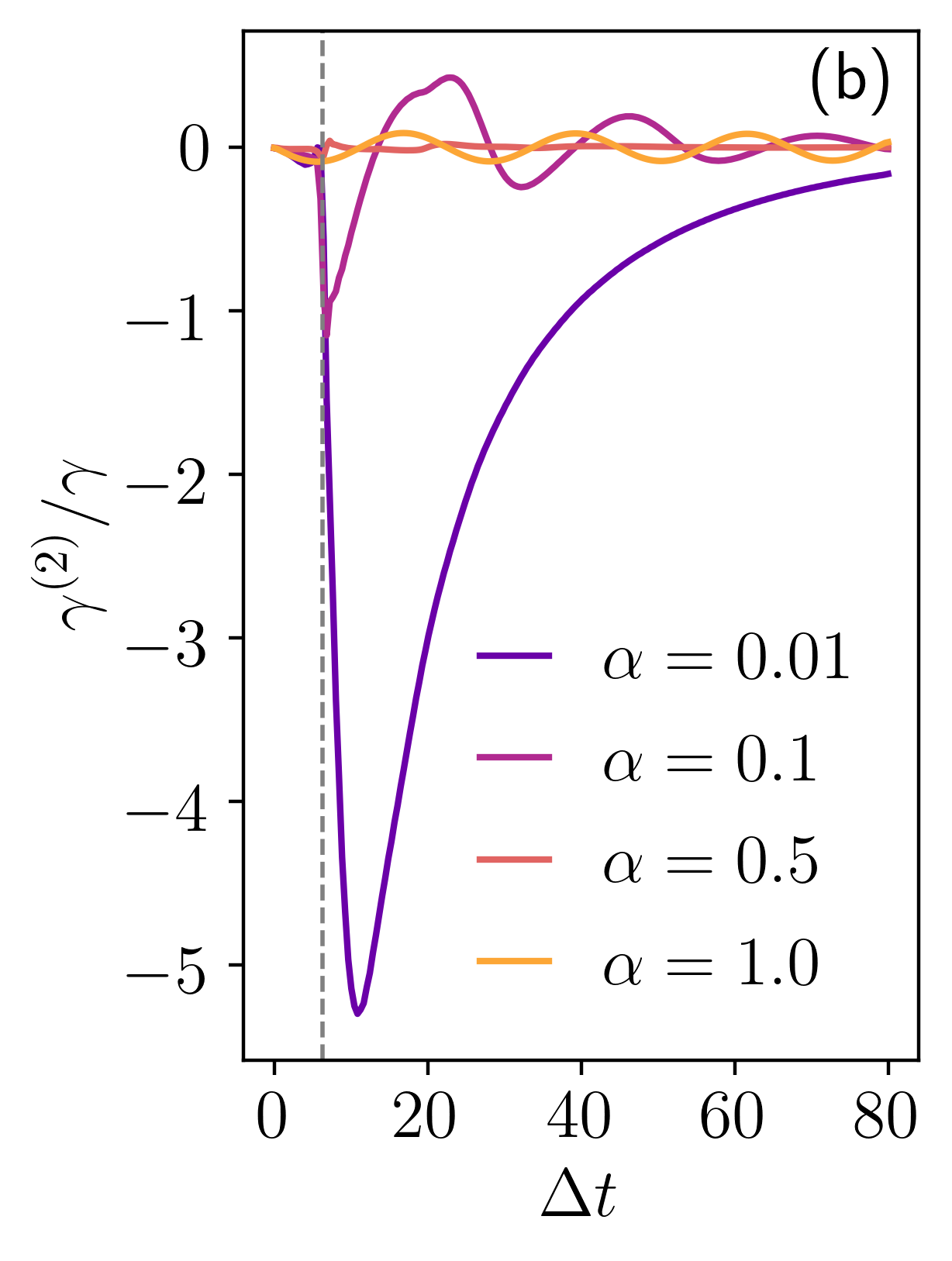}
    \caption{Dynamics of the instantaneous decay corresponding to the first (a) and second emitter (b) in Fig. \ref{fig::eg::state}, for increasing coupling $\alpha$. Other parameters are the same as in Fig. \ref{fig::eg::state}.}
    \label{fig8}
\end{figure}

The analysis of non-Markovian dynamics allows a description in terms of a master equation that is local in time \cite{breuer}. It turns out that the time-dependent generator for this master equation can be written in the standard Lindblad form with time-dependent decay rates. Following this approach, and from our analysis in the polaron picture, we can define a time-dependent decay rate for each emitter as
\begin{eqnarray}
\label{effective::gamma}
\gamma^{(i)}(t)=-2\Re{\left[\frac{\dot c_{i}(t)}{c_{i}(t)}\right]}.
\end{eqnarray}
In general, as in the case of probability amplitudes in the strong coupling regime, an analytical evaluation of this formula is not possible, and we must resort to numerical calculations based on the discrete model. The time evolution of this quantity will contain both sources of non-Markovianity, the one due to the memory of the bath, and the influence of retardation effects. Of course, as  Eq. (\ref{effective::gamma}) is defined in terms of the polaron probability amplitudes, we need to transform it back to the laboratory frame. In Fig. \ref{gamma::tiempo} we show the numerical calculation for the collective dynamics of the time-dependent decay rate for symmetric ($\gamma^{+}$) and antisymmetric ($\gamma^{-}$) initial states. 
{\color{red}

}
As expected, for weak coupling, the decay rate starts to oscillate around the  Fermi Golden rule value (almost without renormalization) $\gamma^{\pm}=\gamma=\pi\alpha\Delta_r \cong \pi\alpha\Delta$ until the feedback from the other emitter suddenly accelerates the decay process, reaching a maximum value of $\sim4.4\gamma$ for the symmetric state in Fig. \ref{gamma::tiempo}(a).  At this point, the emitters see each other, forming a collective state and emitting in a superradiant manner beyond the Markovian Dicke limit of $2\gamma$ for two emitters located at the same position.
This non-Markovian collective enhancement of spontaneous emission has been reported recently within the RWA limit in Ref.  \cite{Kanu19}, where it was shown that larger values can be reached at a particular critical distance. 
We must emphasize that even for weak coupling the Ohmic environment model used here is not flat, so it is reasonable to expect that our calculation of the decay rate does not match exactly the results for a flat environment. Besides, it is clear that non-Markovian processes involving retardation are also present beyond the RWA limits. 
On the other hand, 
for the antisymmetric initial state in Fig. \ref{gamma::tiempo}(b) we have similar behavior at times $t<x/v_g$.  Once they are causally connected, in this case, the antisymmetric state is a Dark state and no decay is observed. 
For weak coupling this is compatible
with the appearance of bound states in the continuum for subradiant initial states in RWA \cite{Tuffarelli2013,Kanu19}. We also have verified
that for slightly distant emitters in the weak coupling we recover the usual Markovian dynamics for initial symmetric (antisymmetric) states.
As the coupling is increased we observe the suppression of emission caused by the qubit renormalization frequency, i.e., $\gamma=\pi\alpha\Delta_r$ accompanied by an oscillating behavior accounting for emission and re-absorption of radiation, which explains the negative values of the decay rate. In the deep-strong coupling regime, where  $\alpha\rightarrow 1$, the system undergoes a localization transition with zero magnetization and the two-qubit system does not radiate anymore.
%
It is worth recalling that what determines the nature of the collective state generated after both qubits become causally connected is the actual phase gained by the radiation field during propagation inside the waveguide. This phase is responsible of the  constructive (destructive) interference of the delay dynamics \cite{Kanu19,Kanu2020}. Within the RWA this phase is given by $\Delta x/ v_g$, which is independent of the qubit-field coupling. In the USC regime the accumulated phase is now coupling and distance-dependent for finite distant emitters as we saw in section \ref{two::qubits::static}, trough the renormalized qubit frequency, dropping to zero in the localized phase for a particular critical coupling. This is a purely USC phenomenon, in which non-Markovian interference effects are also affected by the coupling to the waveguide as we observe in Fig. \ref{gamma::tiempo}. In particular superradiant and subradiant behaviors arise as appropriate combinations of the symmetry of the initial state, the coupling strength, and the relative distance between the emitters, reflecting the complex relation between the Hamiltonian parameters in this regime, which has to be taken into account in real experiments aiming to investigate non-Markovian delay dynamics in waveguide-QED \cite{ferreira2020}.
As complementary information we show in Fig. \ref{fig8} the individual emission (absorption) rates, where the initial state is $\ket{e}\ket{g}\ket{\bold{0}}$ in the polaron picture in correspondence with Fig. \ref{fig::eg::state}, showing the anticorrelated dynamics of both emitters. In this case one can clearly see the influence of Rabi oscillations in the two-qubit system over the spontaneous emission (absorption) rates at strong coupling in the localized regime.  
\\
\section{Summary and Conclusions}
\label{conclusions}
Employing theoretical tools based on the polaron transformation, we have addressed the problem of two distant emitters interacting with a one-dimensional Ohmic waveguide beyond RWA. We have tested early results for the static properties of the two-impurity SBM in the continuum limit, comparing those with a discretized finite $N$-model based on circuit-QED, finding fully agreement for relevant physical quantities. In particular, the induced Ising-like interaction, frequency renormalization, and the modification of the critical coupling for the distance-dependent localization transition in the SBM. We have also shown that it is possible to obtain analytical expressions for the ground state, its magnetization, and the linear entanglement entropy as a function of the relative distance between the emitters. For the study of dynamical properties we have used a time-dependent polaron ansatz, from which we were able to obtain effective polaron non-Markovian dynamical equations in the weak coupling regime accounting for retardation effects of the emitted radiation. We have performed numerical simulations to compare those results with the ones obtained from the proposed discrete model and explore the full dynamical behavior for finite separation distance between the emitters. In the latter case, we have also recovered recent results in the limit of the RWA, and discussed some interesting aspects when entering the so-called ultrastrong coupling regime with finite distant emitters, like the Fermi paradox and the dynamics for particular entangled initial states. We have shown that the dynamics of symmetric (antisymmetric) correlated initial states in the USC regime are strongly affected by non-Markovian delay effects, breaking their indistinguishable exponential decay in the Markovian limit even before qubits are causally connected.
We have also shown that collective dynamics, e.g., \emph{superradiance} or \emph{subradiance}, depend not just on the distance traveled by photons between emitters in the waveguide, but also on the coupling. We remark that this behavior is an exclusive trait of the USC regime that can be potentially exploited for the control of collective states in ultrastrong wQED.
In short, we have established a theory that can be used to study other aspects of time delay or quantum feedback in the USC regime. As a final note, while writing this manuscript, the physics beyond two level emitters in the deep and extremely deep coupling regimes were reported \cite{ashida2021}.  Their treatment is based on a unitary transformation that resembles the Polaron transformation.  In the future, it seems interesting to explore the physics described here in such extreme coupling regimes and  beyond two-level emitters and few excitations.

\begin{acknowledgments}
The authors acknowledge funding from the EU (COST Action 15128 MOLSPIN, QUANTERA SUMO and FET-OPEN Grant 862893 FATMOLS), the Spanish MICINN (MAT2017-88358-C3-1-R), the Gobierno de Arag\'on (Grant E09-17R Q-MAD). 
The authors would also like to thank Kanu Sinha for fruitful discussions on this work.
\end{acknowledgments}
\begin{widetext}
\section{appendix}
\subsection{Ohmic Spin-Boson model in circuit-QED}
\label{circuit::spin::boson}
In this appendix we review the microscopic description of the SB Hamiltonian in the circuit-QED context \cite{GarciaRipollNotes}. For simplicity we treat the case of a single-qubit, but the generalization to $N_q$ qubits is straightforward.
The complete classical Hamiltonian describing  the circuit model is (see Fig. \ref{fig1}(c))
\begin{eqnarray}
\mathcal{H}= \mathcal{H}_q+\mathcal{H}_{L}+\mathcal{H}_{I},
\end{eqnarray}
where each contribution represents the energy of the two-level system or qubit, the transmission line, and their mutual interaction, respectively.
\subsubsection{Qubit Hamiltonian}
For the two-level system we consider a circuit model describing a charge qubit with classical Lagrangian
\begin{eqnarray}
\mathcal{L}_{q}(\Phi,\dot \Phi)=\frac{C_J}{2}\dot\Phi^2+\frac{C_g}{2}\left(\dot\Phi-V_{g}\right)^2+E_{J}\cos\left(\frac{2\pi\Phi}{\Phi_0}\right),
\end{eqnarray}
where $V_g$ is the bias voltage, $C_g$ the coupling capacitance, and $C_J$, $E_J$ the Josephson junction capacitance and energy, respectively. The corresponding Hamiltonian can be simply obtained by a Legendre transformation: $\mathcal{H}_{q}=q\dot\Phi-\mathcal{L}$, being $q=\partial\mathcal{L}_q/\partial\dot\Phi$ the conjugate generalized momentum. The resulting Hamiltonian, after canonical quantization of the coordinates ($[\Phi,q]=\imag$) reads
\begin{eqnarray}
\mathcal{H}_{q}=\frac{(2e)^2}{2 C_{\Sigma}}\left(n-n_g\right)^2-E_{J}\cos\left(\frac{2\pi\Phi}{\Phi_0}\right). 
\end{eqnarray}
Here $C_{\Sigma}=C_{g}+C_{J}$ is the total capacitance, and $n=q/2e$ is the number operator describing Cooper-pairs excitations. The constant term $n_g=C_g V_{g}/2e$ is the gate charge. Writing the above Hamiltonian in the number basis we obtain
\begin{eqnarray}
\mathcal{H}_{q}=4E_{C}\sum_{n}\left(n-n_g\right)^2\proj{n}-\frac{E_{J}}{2}\sum_{n}\left(\ket{n}\bra{n+1}+\rm{h.c.}\right).\nonumber\\
\end{eqnarray}
Truncating the basis to the lowest energy states around the degeneracy point $n_g$, provided that $E_{C}\ll E_{J}$, we obtain the effective qubit Hamiltonian
\begin{eqnarray}
\mathcal{H}_{q}=-\frac{E_{C}}{2}\left(1-2n_{g}\right)\sigma^{z}-\frac{E_{J}}{2}\sigma^{x},
\end{eqnarray}
where the Pauli spin operators are defined in the charge basis. 
\subsubsection{Discrete Transmission line}
\label{discrete::model}
For the transmission line we follow a similar procedure by considering a discretized line of length $L$ with $N$ nodes, such that the discretization distance is $\Delta x=L/N$ (see Fig. \ref{fig1}(c)). We define $C_0$ and $L_0$ as being the characteristic capacitances and inductances of the lumped circuit model, respectively. The Lagrangian in this case is written as
\begin{equation}
    \mathcal{L}_{L} =\frac{1}{2} \sum_{n=1}^{N}\left[ C_{0}\dot\Phi_{n}^2-\frac{\left(\Phi_{n+1}-\Phi_{n}\right)^2}{ L_0}\right].
\end{equation}
The corresponding conjugate momentum (charge coordinate) is given by $q_{n}$=$\partial\mathcal{L}_L/\partial{\dot\Phi_{n}}$=$ C_0\dot\Phi_{n}$. 
The Hamiltonian is obtained by applying the Legendre transform: $\mathcal{H}_{L}$=$\sum_{n}q_{n}\dot\Phi_{n}-\mathcal{L}_L$, which results in
\begin{eqnarray}
\label{Hamiltonian::circuit::model}
\mathcal{H}_{L} =\frac{1}{2} \sum_{n=1}^{N}\left[\frac{q_{n}^2}{C_0}+\frac{\left(\Phi_{n+1}-\Phi_{n}\right)^2}{ L_{0}}\right].
\end{eqnarray}
This Hamiltonian is identical to that of a linear chain of classical $N$ coupled harmonic oscillators, therefore it can be diagonalized by Fourier-transforming to the momentum space. Here we impose periodic boundary conditions such that $\Phi_{1}\equiv\Phi_{N+1}$, hence $k_{m}=2\pi m/L$. The diagonal Hamiltonian can be written as
\begin{eqnarray}
\mathcal{H}_{L}=\frac{1}{2}\sum_{k}\left[\frac{q_{k}q^{*}_{k}}{C_0}+\left[2-2\cos\left(2\pi m/N\right)\right]\frac{\Phi_{k}\Phi^{*}_{k}}{ L_0}\right],
\end{eqnarray}
where $\Phi_{k}=N^{-1/2}\sum_{n}\exp\left(\imag k x_{n}\right)\Phi_{n}$ is the magnetic flux in momentum space. This leads to the dispersion relation of the transmission line
\begin{eqnarray}
\omega_{k}&=&\frac{1}{\sqrt{C_0 L_0}}\sqrt{2-2\cos\left(k_m\Delta x\right)} \nonumber\\
&=&\omega_c\sqrt{2-2\cos\left(k_m\Delta x\right)},
\end{eqnarray}
where we have defined the cutoff frequency $\omega_c=v_{g}/\Delta x$, being $v_g=1/\sqrt{c_0 l_0}$ the group velocity of photons in the transmission line.
In the continuum limit, i.e., $N\xrightarrow{}\infty$ (or $\Delta x\xrightarrow{}0$) we have that
\begin{eqnarray}
\label{continious::Hamiltonian}
\mathcal{H}_L=\frac{\Delta x}{2}
\sum_{k}\left[\frac{\tilde q_{k}\tilde q^{*}_{k}}{c_0}+\left(\frac{2\pi m}{N}\right)^2\frac{\Phi_{k}\Phi^{*}_{k}}{l_0}\right],
\end{eqnarray}
where $\tilde q_{k}=q_{k}/\Delta x$ is the charge density. From the continuous Hamiltonian in Eq. (\ref{continious::Hamiltonian}) we find the well known linear dispersion relation
\begin{eqnarray}
\label{linear::dispersion}
\omega_k\approx v_{g}|k|.
\end{eqnarray}
Fig. \ref{fig_wk} shows the dispersion relation for the discrete transmission line and its approximate linear behavior for low momentum. 
\begin{figure}
    \centering
    \includegraphics[scale=0.5]{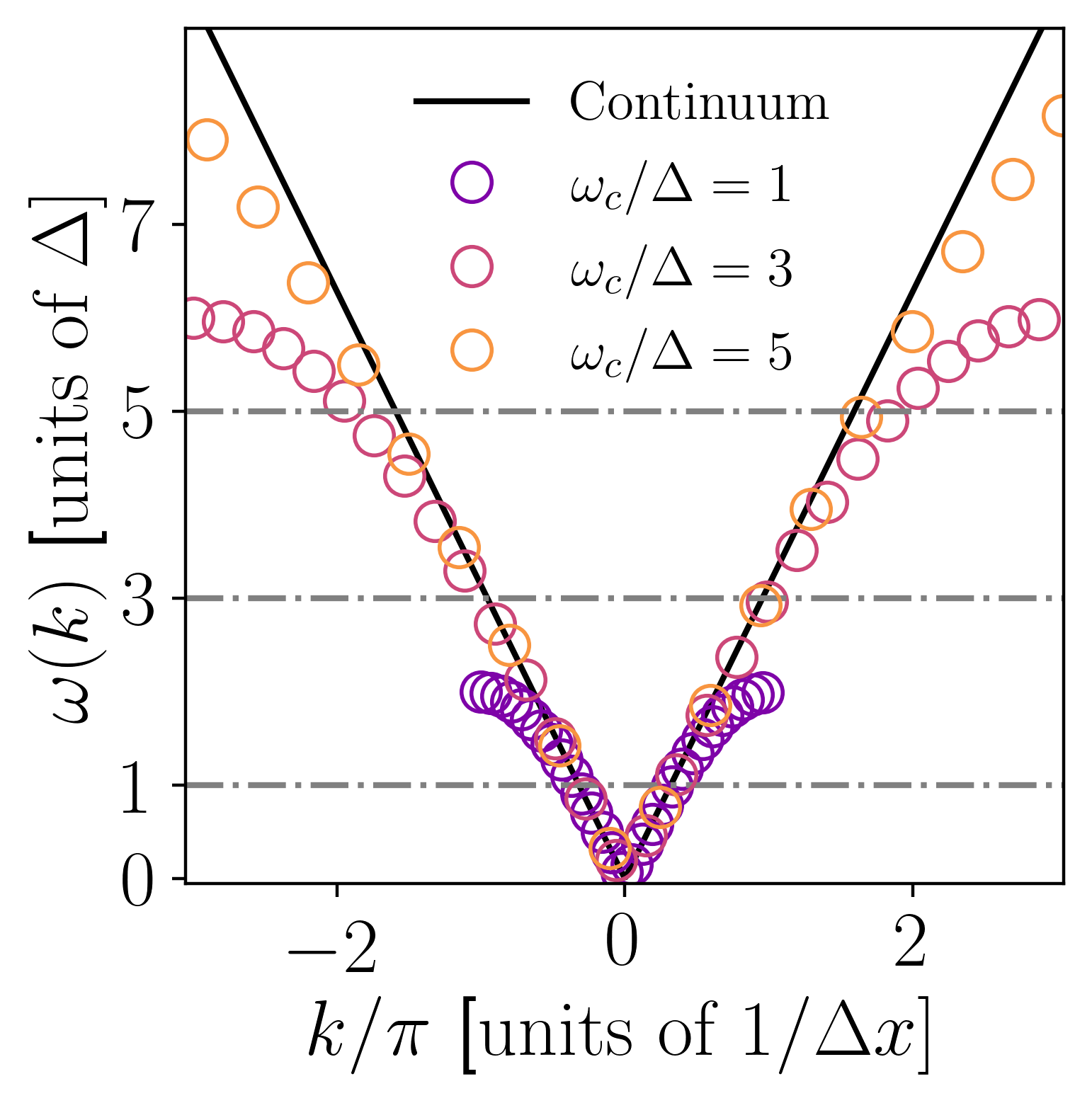}
        \caption{Dispersion relation of the transmission line with periodic boundary conditions. Empty circles indicate results for the discrete model and the solid line shows the linear behavior in the continuum limit. Three cases for different cutoff frequencies (dashed-dotted lines) are shown : $\omega_c/\Delta=\{1.0, 3.0, 5.0\}$.}
    \label{fig_wk}
\end{figure}
\subsubsection{Microscopic Interaction}
The capacitive interaction between the superconducting qubit and the discrete transmission line is given by
\begin{eqnarray}
\mathcal{H}_{I}=e\frac{C_{g}}{C_{\Sigma}}\sigma^{z}V(x). 
\end{eqnarray}
By using the fact $V(x)=\partial_{t}\Phi=-\imag\left[\Phi,\mathcal{H}\right]$, where the quantum flux field for the transmission line satisfies the one-dimensional massless Klein-Gordon scalar equation, whose solution reads \cite{Tong2007}
\begin{eqnarray}
\Phi(x) =\sum_{k}\sqrt{\frac{1}{2c_{0}\omega_{k}L}}\left(a_{k}e^{-\imag k x}+\rm{h.c.}\right),
\end{eqnarray}
we arrive to 
\begin{eqnarray}
\mathcal{H}_{I}=\sigma^{z}\sum_{k}\left(g_{k}a_{k}e^{-\imag k x}+\rm{h.c.}\right),
\end{eqnarray}
where 
\begin{eqnarray}
g_{k}=g\sqrt{\frac{\omega_k}{2L}}, \,\,\,\, g=\imag e \frac{C_g}{C_\Sigma}\sqrt{\frac{1}{c_0}}.
\end{eqnarray}
In view of the above discussion we can conclude that the total quantum Hamiltonian reads
\begin{eqnarray}
\mathcal{H}=-\frac{E_C}{2}\left(1-2n^{\rm cl}_{g}\right)\sigma^{z}-\frac{E_{J}}{2}\sigma^{x}+\sum_{k}\omega_k a^{\dagger}_{k}a_{k}+\sigma^{z}\sum_{k}\left(g_{k}a_{k}e^{-\imag k x}+\rm{h.c.}\right),
\end{eqnarray}
which is the general one-qubit SBM up to a unitary rotation of the spin basis \cite{Leggett87}.
\subsubsection{Ohmic Spectral Function}
\label{alphaOhmic}
Once we have calculated the couplings from the microscopic interaction, we can compute analytically the spectral function of the transmission line, i.e., the environment spectral density in the continuum limit, which contains all the information about the bath modes \cite{weiss2012quantum,GarciaRipollNotes}. It reads
\begin{eqnarray}
\label{spectra::function}
J(\omega)&=&2\pi\sum_k |g_k|^2\delta(\omega-\omega_k) \nonumber\\
&=&\frac{|g|^{2}}{v_{g}}\int_{0}^{\omega_c}d\omega_k \omega_k \delta(\omega-\omega_k) \nonumber \\
&=&\frac{|g|^{2}}{v_g}\omega=\pi\alpha\omega, \,\,\,\, \alpha\equiv|g|^{2}/\pi v_{g}.
\end{eqnarray}
As we showed in Fig. \ref{fig1} this Ohmic spectral function can be well approximated with the aforementioned discrete transmission line model. 
Notice that in deriving the above expression we have made use of the linear dispersion relation in the continuum (\ref{linear::dispersion}).
\subsection{Correction terms for Ising Interaction}
\label{Ising::long}
Small corrections can be obtained for the approximate Ising coupling given in Eq. (\ref{J::sinc}) by replacing the complete expression for the variational $f_k$ into Eq. (\ref{Ising::ij}),
\begin{eqnarray}
\mathcal{J}&=&
2\alpha\int_{0}^{\omega_c}d\omega \frac{\omega}{(\Delta_r+\omega)}\cos\left(\omega x/v_g\right)-\alpha\int_{0}^{\omega_c}d\omega\frac{\omega^2}{(\Delta_r +\omega)^2}\cos\left(\omega x/v_g\right).\nonumber\\
&=& \alpha\omega_c{\rm{sinc}}(\omega_c x/v_g)+\frac{\alpha\Delta_r^2\cos(\omega_c x/v_g)}{\delta_r}\nonumber\\
&+&\alpha\Delta_r\left[\frac{x\Delta_r}{v_g}\sin(x\Delta_r/v_{g})\left[{\rm{Ci}}(x\Delta_r/v_g)-{\rm{Ci}}(x\delta_r/v_g)\right]\right. \nonumber\\
&+& \left.\frac{x\Delta_r}{v_g}\cos(x\Delta_r/v_{g})\left[{\rm{Si}}(x\delta_r/v_g)-{\rm{Si}}(x\Delta_r/v_g)\right]-1\right],
\end{eqnarray}
with $\delta_r=\Delta_r+\omega_c$, and ${\rm{Ci}}(x)=-\int_{x}^{\infty}dt \cos t/t$, ${\rm{Si}}(x)=\int_{0}^{x}dt \sin t/t$.
\subsection{Reduced density matrix and linear entropy}
\label{entropy::large}
Tracing over the bath degrees of freedom we can get the two-qubit ground state density matrix in the basis $\{\ket{--},\ket{-+},\ket{+-},\ket{++}\}$:
\begin{eqnarray}
\varrho_{\rm {GS}}=\frac{1}{2}
\left[\begin{array}{cccc}
\frac{1}{2}+\sin\theta\cos\theta & -\frac{\Delta_r}{2\Delta}\cos2\theta & -\frac{\Delta_r}{2\Delta}\cos2\theta e^{\imag\phi(x)} & \frac{\Delta_r^2}{\Delta^2}\left(\frac{1}{2}+\sin\theta\cos\theta\right)e^{-\zeta(x)}	\\
-\frac{\Delta_r}{2\Delta}\cos2\theta & \frac{1}{2}-\sin\theta\cos\theta & \frac{\Delta_r^2}{\Delta^2}\left(\frac{1}{2}-\sin\theta\cos\theta\right)e^{\zeta(x)} & -\frac{\Delta_r}{2\Delta}\cos2\theta e^{-\imag\phi(x)}	\\
-\frac{\Delta_r}{2\Delta}\cos2\theta e^{-\imag\phi(x)} & \frac{\Delta_r^2}{\Delta^2}\left(\frac{1}{2}-\sin\theta\cos\theta\right)e^{\zeta(x)} & \frac{1}{2}-\sin\theta\cos\theta & -\frac{\Delta_r}{2\Delta}\cos2\theta	\\
\frac{\Delta_r^2}{\Delta^2}\left(\frac{1}{2}+\sin\theta\cos\theta\right)e^{-\zeta(x)} & -\frac{\Delta_r}{2\Delta}\cos2\theta e^{\imag\phi(x)} & -\frac{\Delta_r}{2\Delta}\cos2\theta & \frac{1}{2}+\sin\theta\cos\theta
\end{array}\right], \nonumber\\
\end{eqnarray}
where $\phi(x)=4\sum_{k}f^{2}_{k}\sin k x$, and $\zeta(x)=4\sum_{k}f^{2}_{k}\cos k x$.
Using the above density matrix, a general expression for the linear entanglement entropy reads 
\begin{eqnarray}
\label{entropy::general}
S_L&=&1-\rm{Tr}\varrho_{\rm GS}^2\nonumber\\
&=&1-\frac{\Delta_r^2 \cos ^2 2 \theta}{2 \Delta ^2}-\frac{1}{2} \left(\frac{1}{2}-\sin\theta\cos\theta
\right)^2-\frac{1}{2} \left(\sin\theta \cos \theta+\frac{1}{2}\right)^2 \nonumber\\
&-&\frac{\Delta_r^4 e^{2 \zeta(x)}
\left(\frac{1}{2}-\sin\theta\cos\theta \right)^2}{2 \Delta ^4}-\frac{\Delta_r^4 e^{-2\zeta (x)}
\left(\sin\theta\cos\theta +\frac{1}{2}\right)^2}{2 \Delta ^4}.\nonumber\\
\end{eqnarray}
\end{widetext}
\bibliography{ref}
\end{document}